\documentclass[twocolumn, 5p]{elsarticle}

\usepackage{amsmath, amssymb}
\usepackage{breqn}

\usepackage{color}
\usepackage{graphicx}
\usepackage{dcolumn}
\usepackage{bm}
\usepackage[all]{xy}

\newcommand{\beq}{\begin{equation}}
\newcommand{\eeq}{\end{equation}}

\begin{document}

\title{Stochastic Dynamics of Extended Objects in Driven Systems:\\ I. 
Higher-Dimensional Currents in the Continuous Setting}
\author[mjc]{Michael~J.~Catanzaro}
\author[mjc,vyc]{Vladimir~Y.~Chernyak}
\author[mjc]{John~R.~Klein}

\address[mjc]{Department of Mathematics, Wayne State University,
656 W. Kirby, Detroit, MI 48202}

\address[vyc]{Department of Chemistry, Wayne State University,
5101 Cass Ave, Detroit, MI 48202}




\date{\today}

\begin{abstract}
The probability distributions, as well as the mean values of stochastic
currents and fluxes, associated with a driven Langevin process, provide a good
and topologically protected measure of how far a stochastic system is driven
out of equilibrium. By viewing a Langevin process on a compact oriented
manifold of arbitrary dimension $m$ as a theory of a random vector field
associated with the environment, we are able to consider stochastic motion of
higher-dimensional objects, which allow new observables, called
higher-dimensional currents, to be introduced. These higher dimensional currents
arise by counting intersections of a
$k$-dimensional trajectory, produced by a evolving $(k-1)$-dimensional cycle, with a
reference cross section, represented by a cycle of complimentary dimension
$(m-k)$. We further express the mean fluxes in terms of the solutions of the
Supersymmetric Fokker-Planck (SFP), thus generalizing the corresponding
well-known expressions for the conventional currents. 
\end{abstract}

\maketitle

\section{Introduction}
\label{sec:intro}

Stochastic Langevin processes appear in a wide variety of 
fields~\cite{CKW96, LWDZ03, SN07, CBL15, G00, O13}. This is not
surprising if one thinks of these as a result of elimination of fast
environmental (bath) degrees of freedom, resulting in dissipative and noise
terms in the reduced equations for the system~\cite{LL02, CKW96}. The popularity of the
Langevin approach in describing dynamical systems that occur in different
applications can be explained in a variety of ways. The Langevin equation is
simple, universal, and adequately describes the classical dynamics of an open system
whose dynamical time scale is slow compared to the bath relaxation time,
including the system relaxation to its steady state~\cite{R84, CKW96}. In the case of
detailed balance, the steady state is represented by a Boltzmann 
distribution~\cite{VK92}.

Stationary systems whose steady state is a true equilibrium, characterized by
the absence of fluxes of any kind, as of today can be described as well as
studied, at least on the conceptual level, in both the classical and quantum 
cases~\cite{R09}.
Steady states of generic or, in other words, driven systems, often referred to
as non-equilibrium steady states, are much less understood, even in the case of
classical open systems coupled to a fast bath, when the latter degrees of
freedom can be efficiently eliminated, resulting in a Langevin equation~\cite{D07}.
Despite the fact that Langevin processes can be efficiently studied, e.g., by
switching from the Lagrangian language of stochastic trajectories to the
Euler-Hamilton picture of deterministic Fokker-Planck (FP) equations for the
probability distributions~\cite{CKW96, G00, VK92, R84}, theories for the steady states of driven systems are
still in the developmental stage. This is not surprising since, even in the
absence of noise, a driven system is described by a generic system of ODEs and
its behavior can be very rich, ranging from integrable systems to systems which
possess various degrees of chaos. On the other hand, studying non-equilibrium
systems is really necessary, since most systems of interest are in fact driven,
e.g., biological systems~\cite{AA07, A07, JAP97, LWDZ03, NYYK97}, environmental 
systems~\cite{P98}, as well as various kinds of
networks, including electrical power grids~\cite{CBL15}, 
gas supply networks~\cite{CBL15, CFBBM15}, 
networks describing price markets~\cite{B98, LOT02}, and many others~\cite{RHJ09, MWP14, AA07, P98}.

Over the past three decades a substantial effort has been put into not only studying
the behavior of specific driven systems, but also in identifying a set of basic
principles and laws that would govern them.
In other words, such efforts amount to building  
an analogue of thermodynamics, often referred to as non-equilibrium 
thermodynamics~\cite{DGM13, J98, HS01}.
A large variety of such laws, universal exact
relations, and concepts have already been identified. These include various
kinds of fluctuation theorems for produced work~\cite{AG07, HJ07, CCKP04}, 
generated heat and entropy~\cite{GC95, C99, C98},
and similar observables, as well as their reduced counterparts, known as the
Jarzynski relations~\cite{J97a, J97b, H99, J98, CCJ05}.

A very useful concept that has been drawing more and more attention recently is
that of a stochastic current, or 
flux~\cite{SN07, RHJ09, O08, D07, BDGJL06, CCBK13}. It would be worth
 emphasizing, that although stochastic current densities appear in any driven
system, to have a non-trivial flux the configuration space of a system should
have non-trivial topology~\cite{CCMT09}. In fact, it should have non-contractible
$1$-dimensional cycles; in other words, flux has a topological nature. 
The topological nature of flux implies it is a topologically protected quantity,
since it is based on counting the total number of events
by the total time to obtain the rates. As outlined, e.g., in \cite{CCMT09}, a
flux can be measured in two equivalent ways, by counting some kind of a winding
number, i.e., how many times the system spans a given non-contractible cycle,
or alternatively, by how many times a stochastic trajectory crosses a reference
cross section. The equivalence between the two approaches to a measurement was
established in~\cite{CCMT09}, and relies on one of the most basic principles in
algebraic topology of manifolds, known as Poincar\'e
duality~\cite{Spanier95,GH94}. Currents and fluxes are also well-defined for
the discrete counterpart of Langevin processes, namely continuous-time Markov
Chain (MC) stochastic processes on graphs~\cite{CKS12a,CKS13}. In this case,
current/flux is measured by counting how many times a stochastic trajectory
(Markov chain) goes over a given graph edge, with the proper sign that accounts
for the direction.  There is another good reason for the importance of the
current/flux concept: it turns out that the probability density together with
current density constitute the right set of variables for studying the
long-time limit of the probability distributions. The large-deviation
(Cram\'er) function ${\cal S}(\rho, \bm{J})$, referred to as the universal
current density functional that describes the long-time behavior of the
probability distribution function (functional) of the density $\rho$ and
current density $\bm{J}$, has a simple form and can be written explicitly in
the continuous as well as discrete cases, see, e.g.~\cite{BD04, BDGJL05, 
BDGJL06, GKP06, D07, MN08, MNW08, CCMT09}.
This fact forms the basis of a field,
often referred to as the $2.5$-level theory~\cite{BC15}.  The $(\rho, \bm{J})$
description turned out to also be important in stochastic optimal control
theory, leading to the so-called Gauge-Invariant-Hamilton-Jacobi-Bellman (HJB)
equation that extends control to cost functionals with the terms linear in
velocity, as well as providing an optimization view of the celebrated HJB
equation~\cite{CCBK13,BCCK16}.

On the one hand, studying the probability distributions of currents and fluxes
in specific systems has shown the usefulness of the concept by providing
non-trivial measures of the driven nature of the underlying stochastic
dynamics. On the other hand, a large number of exact relations for generated
currents in driven systems has been established, so far mostly in the discrete
setting, thus making the current/flux good candidates for building a theory
of non-equilibrium thermodynamics. While the fluctuation theorems for entropy
production and related quantities~\cite{C99, C00, K98, HS07, HV10, S05} 
are despite their
non-triviality,  well understood by now, and in most cases follow from
comparing the probability of a stochastic trajectory with its time-reversed
counterpart, the exact relations for currents and fluxes in both
open~\cite{CCS11} and closed~\cite{SAC11} networks, turns out to be more
surprising. A very interesting class of effects is related to the situation of
periodic driving, when at any given time the system is in detailed balance,
while the system parameters change in time in a periodic fashion, with the
latter dependence being the only source of driving~\cite{CS10, MM01}. For this
setting a variety of exact statements have been identified, including
no-pumping and pumping restriction statements~\cite{CKS12a, CS08}.  One of the
most interesting findings was realizing that the fluxes are still generated in
the adiabatic limit, when effects of driving become minimal, and have geometric
nature, being interpreted in terms of stochastic Berry phases~\cite{O08, S09,
SN07}. Further studies of the adiabatic low-temperature limit revealed the flux
quantization effect~\cite{CKS12a, CKS12b, CKS13}, i.e, in this limit the flux,
defined as the number of counted events per driving protocol becomes an 
integer in a generic case, and rational in the presence of permanent degeneracies; the
effect has been observed experimentally.

In this manuscript, we generalize the notion of a stochastic current/flux to a
higher-dimensional case, thus providing new observables associated with
Langevin processes in continuous spaces, that characterize in a robust, topologically protected way the extent to which a stochastic system is driven out of equilibrium. In other words we bring in topological concepts that provide a better understanding of physics of non-equilibrium phenomena, as well as new insights to non-equilibrium thermodynamics.
The aforementioned generalization is very
natural and in fact simple. Starting with the interpretation of a stochastic
flux as an intersection index (the sum over intersection points weighted with
the proper $\pm 1$ sign factors to account for direction) of a stochastic
trajectory in a space $X$ of dimension ${\rm dim}(X) = m$ with a cross section,
we extend our consideration to stochastic motion of higher-dimensional objects.
Namely, we consider $(k-1)$-dimensional cycles that span $k$-dimensional trajectories and
count their intersections with a reference cross section, represented by a
cycle of complimentary dimension $(m - k)$, weighting them with the proper sign
factors, resulting in the intersection index of two manifolds (a stochastic
trajectory and the cross section). Summarizing, the higher-dimensional flux is
an observable that associates with a $k$-dimensional stochastic trajectory its
intersection index with a reference $(m-k)$-dimensional cross section per unit
time. To define higher-dimensional trajectories generated in a Langevin
process, one needs to do a very simple and natural thing of replacing the
traditional correlation function of the random field in
Eq.~\eqref{eq:Langevin-standard} with a more general one in
Eq.~\eqref{eq:Langevin-general}, which immediately interprets a Langevin process
as what it actually is---a theory of random vector fields, or equivalently, a
theory of stochastic flows.

In the case of standard currents, the observables can be efficiently computed by
switching from the Lagrangian language of stochastic trajectories to the
Euler-Hamilton picture of deterministic linear PDEs for the probability
distributions~\cite{CKS12a, CKS12b, SN07}. The Euler-Hamilton picture
involves studying the Fokker-Planck (FP; or Kolmogorov in the mathematical
literature) equations. To generalize the FP-equation approach to
higher-dimensional fluxes, we adopt the proposal of Tanase-Nicola and Kurchan
\cite{T-NK04}, who showed on an intuitive level that the so-called
Supersymmetric Fokker-Planck (SFP) equation adequately describes stochastic
evolution of higher-dimensional objects. In fact, we formalize the ideas
presented in \cite{T-NK04} to develop an interpretation of the super-states
$\varrho(x^1, \ldots, x^m; \Theta^1, \dots, \Theta^m)$ that depend not only on
the standard set coordinates $x$, but also on a set of Grassmann (anticommuting)
variables, and satisfy the SFP equation, as {\it reduced measures} that contain
reduced, yet all the necessary information on the complete probability
distributions $d{\cal P}(\eta)$ in the infinite-dimensional (functional) of
$(k-1)$-dimensional cycles in $X$, with the reduction being compatible with
stochastic evolution. The above interpretation made it possible to obtain the
main result of this manuscript for continuous models--closed expressions for
average higher-dimensional stochastic fluxes in terms of the solutions of the
SFP equations for the general time-dependent, stationary and periodic-driving
cases.

We would like to emphasize that the higher-dimensional fluxes are not
associated with a new class of stochastic models, but rather provide a set of
new topologically protected observables, associated with the same ``good old''
Langevin processes. However, they give rise to a new class of discrete models,
defined on CW-complexes, which are higher-dimensional generalizations of
graphs. These more general discrete models, studied in the second manuscript,
appear in a natural way
by considering the long-time relaxation of higher-dimensional cycles in the
low-temperature limit for a Langevin process, in the same way as a Markov
process on a graph can be interpreted as slow relaxation between the local
minima of the potential function $V(x)$, represented by the graph vertices, via
rare over-the-barrier transition events, with the transition paths represented
by the graph edges; actually describing a multi-state thermo-activated chemical
reaction, considered within the multi-dimensional version of the celebrated
Kramers transition state theory~\cite{K40,HTB90}. After deriving the CW-complex models,
including evolution equations for the reduced measures (a discrete counterpart
of the reduced measures for Langevin processes), we focus on the periodic driving
case. The main result of the second manuscript 
proves there is an explicit formula for the generated flux in
the adiabatic limit, which generalizes the expression for a Markov process on a
graph, obtained in our earlier work \cite{CKS12a}, \cite{CKS12b}, \cite{CKS13}, 
to the higher-dimensional case. To that
end, we express the two ingredients of this expression, namely the solution of
the higher-dimensional Kirchhoff network problem~\cite{CCK15a} and 
the higher-dimensional Boltzmann distribution~\cite{CCK15b}, 
as a weighted sum over spanning trees
and co-trees. These results are referred to as the higher-dimensional Kirchhoff tree
and co-tree theorems, respectively. The aforementioned expression allows us
to demonstrate that
in the low-temperature, adiabatic limit, the generated fluxes are quantized
(pumping quantization theorem). However in a generic case quantization is
rational, rather than integer, which reflects higher complexity in the
stochastic evolution of higher-dimensional cycles.

At this point we would like to note that although the dynamics we are dealing
with in this manuscript is a standard Langevin process that is a common tool in
chemical physics, the topological and higher-dimensional nature of the new
observables requires a certain number of theoretical techniques, which are
common in topology and geometry, rather than in chemical physics. Still, the
problems addressed in this manuscript belong to the scope of chemical physics.
Therefore, to make the manuscript accessible for the chemical physics community
we have put substantial effort to describe all the concepts and techniques,
involved in the derivations at least on an intuitive level, and with enough
detail, so that a reader can follow the derivations without a necessity to read
additional mathematical literature.

This manuscript is organized as follows. In section~\ref{sec:tech-intro}, we
describe the main concepts and technical tools involved in our derivations,
which are not very traditional in the chemical physics
community, as well as formulate the main results presented in these two manuscripts.
A detailed outline of the material presented there is given at the beginning of
the section; here we just note that the main results of this manuscript for
higher-dimensional fluxes in continuous stochastic systems are
presented in subsections~\ref{sec:higher-flux-driven}. 
Section~\ref{sec:Langevin-current-supersymmetry} is
devoted to the derivations of our main results in the continuous setting,
formulated in subsection~\ref{sec:higher-flux-driven}. In
subsection~\ref{sec:generating-functions} we formulate a path-integral
representation for the average flux and the generating function. In
subsection~\ref{sec:SFP-currents}, we introduce our main computational tool,
namely the reduced measures, and apply it to convert the path-integral
representations to the Hamilton-Euler approach of fermionic super-states and SFP
equations. In subsection~\ref{sec:current-continuous-explicit}, we provide the
derivations of the final expressions for the average flux in the stationary and
periodically driven cases. In section~\ref{sec:discusssion}, we summarize our results on the continuous setting and discuss some future possible developments/applications.

\section{Technical Introduction}
\label{sec:tech-intro}

In this section, we describe the scope of this work and formulate the main results
presented in the manuscript. Since the aim of this manuscript is rather
technical, we will summarize here the main results and main
concepts needed to formulate the aforementioned results.

The material in this section is organized as follows. 
Subsection~\ref{sec:Langevin-regular} presents a very important
interpretation of Langevin processes as a theory of stochastic flows
generated by random vector fields. Equivalently, Langevin processes
can be thought of as random walks in the space
of diffeomorphisms of the configuration space $X$. In
subsection~\ref{sec:currents}, we introduce the concept of a higher dimensional
cycle $\gamma$, and place an equivalence relation on the cycles to form 
the homology groups $H(X)$, which turn out to be  abelian groups or vector spaces.
Furthermore, we introduce a higher-dimensional
stochastic flux as an intersection index of a stochastic trajectory $\eta$
with a reference cross section $\alpha$, describe the flux in terms of
homology, and introduce the closely related concept of
Poincar\'e duality.  To provide some intuition and insight for higher dimensional
fluxes, we present several simple examples of deterministic flows that generate
non-zero higher dimensional fluxes in
subsection~\ref{sec:higher-currents-intro}. In
subsection~\ref{sec:SFP-ferm-diff-forms}, we provide the necessary facts about 
supersymmetric stochastic dynamics, including the Supersymmetric Fokker-Planck
(SFP) equation, related geometric structures, and de Rham cohomology $H_{{\rm
DR}}(X)$ - the residence of the equivalence classes $[\psi]$ of the fermion
(super)states $\psi$. In section~\ref{sec:higher-flux-driven}, we revisit
Poincar\'e duality and present its formulation as an equivalence between homology
and cohomology in complimentary dimensions. This allows us to formulate and
present the main results of this manuscript,
related to higher-dimensional currents in the continuous setting. Namely, we
provide closed
expressions for average values of stochastic fluxes in terms of solutions of
the SFP equation, including the cases of general, stationary, and periodic potential
driving. 

\subsection{Langevin Processes as Random Walks in Diffeomorphism Groups and Regularization}
\label{sec:Langevin-regular}

We reiterate that we are dealing with Langevin processes on a smooth oriented
compact manifold $X$ of dimension ${\rm dim}\, X=m$. The compactness requirement
can be relaxed, when necessary, at the cost of adding some technical details.

As described above, standard empirical currents can be viewed as a topological
observable for a Langevin process in a manifold $X$ that associates to each
stochastic trajectory its homology class. A Langevin process with Gaussian
Markovian noise can be represented by a stochastic equation
\begin{gather}
  \begin{gathered}
  \dot{x}^j=u^j(x,t)+\xi^j(x,t), \\
  \langle \xi^{j}(x,t) \rangle=0,  \\
 \langle \xi^{i}(x,t)\xi^{j}(y,t') \rangle = 2\kappa G^{ij}(x,y)\delta(t-t'), 
\label{eq:Langevin-general}
\end{gathered}
\end{gather}
where $\xi(t)$ is random vector field with Gaussian Markovian statistics, fully
described by the correlation function $G$, and
$\kappa=\beta^{-1}=(k_{B}T)^{-1}$ controls the noise strength. The tensor field
$g^{ij}(x)=G^{ij}(x,x)$ defines a non-negative metric in $X$ (strictly speaking
a scalar product in the cotangent bundle). We will assume that $g$ is
non-degenerate, which implies that it is positive definite. This means that our
configuration space $X$ is equipped with a Riemannian metric $g_{ij}(x)$, defined
by the condition $g_{ik}(x)g^{kj}(x)=\delta_{i}^{j}$, where $\delta_{i}^{j}$ is
the Kronecker delta, and hereafter we imply summation over repeating
indices. At this point, it is worth noting that in describing a
Langevin process Eq.~(\ref{eq:Langevin-general}) is usually replaced with
\begin{gather}
 \begin{gathered}
  \dot{x}=u(x,t)+\xi(x,t), \\
  \langle \xi^{j}(x,t) \rangle=0, \\
 \langle \xi^{i}(x,t)\xi^{j}(x,t') \rangle = 2\kappa g^{ij}(x)\delta(t-t'), 
\label{eq:Langevin-standard}
\end{gathered}
\end{gather}
which is consistent with Eq.~(\ref{eq:Langevin-general}). The reason is that in a
standard set-up a Langevin process is studied on the level of stochastic
dynamics of points in the configuration space $X$. Therefore, due to the
Markovian, i.e., the $\delta$-correlated in time, nature of a Langevin process,
the correlation properties of the random vector field $\xi(x)$ at the same
point of the configuration space $X$ only are relevant for the standard set-up.
In what follows we will be studying the evolution of higher-dimensional
objects, so that the complete information on the underlying Langevin process,
given by Eq.~(\ref{eq:Langevin-general}) is required. Stated differently, this
implies that the stochastic dynamics of higher-dimensional objects contains
more detailed information on the underlying Langevin process.

Most importantly, the r.h.s. of Eq.~\eqref{eq:Langevin-general} determines a flow
and, therefore, a random Markovian walk in the group ${\rm Diff}(X)$ of
diffeomorphisms of $X$. This should be understood as follows. Vector fields on
$X$ can be naturally interpreted as infinitesimal diffeomorphisms, i.e., the
Lie algebra ${\rm Vect}(X)$ of vector fields in $X$ can be viewed as the Lie
algebra $\mathfrak{a}({\rm Diff}(X))$ associated with the group ${\rm Diff}(X)$
of diffeomorphisms of $X$. Furthermore, since vector fields represent first-order differential operators, a two-fold tensor product of vector
fields represents a second-order differential operator.
The deterministic (advection term) vector field
$u\in {\rm Vect}(X)$ can be considered as the element of the relevant algebra
${\rm Vect}(X)=\mathfrak{a}({\rm Diff}(X))$, whereas we can treat the noise
correlation function as a symmetric element $G\in {\rm Vect}(X)\otimes {\rm
Vect}(X)$. 
This allows the generalized Fokker-Planck (FP) operator
\begin{equation}
\label{FP-general} \hat{{\cal L}}=\kappa G + u
\end{equation}
to be introduced as an element of the universal enveloping algebra $U{\rm
Vect}(X)=U\mathfrak{a}({\rm Diff}(X))$ of the Lie algebra ${\rm Vect}(X)$ of
vector fields in $X$.\footnote{The universal enveloping algebra $U\mathfrak{h}$
of a Lie algebra $\mathfrak{h}$ is
$T\mathfrak{h}=\sum_{p=0}^{\infty}\mathfrak{h}^{\otimes p}=\mathbb{R}\oplus
\mathfrak{h} \oplus (\mathfrak{h}\otimes \mathfrak{h}) \oplus \ldots$ with the
relations $a\otimes b-b\otimes a-[a,b]=0$ for $a,b\in\mathfrak{h}$
\cite{Serre65}. The natural morphism $T\mathfrak{h}\to U\mathfrak{h}$ generates
maps $\mathfrak{h}\to U\mathfrak{h}$ and $\mathfrak{h}\otimes\mathfrak{h}\to
U\mathfrak{h}$ so that with a minimal abuse of notation we can view
$\mathfrak{h}\subset U\mathfrak{h}$ and $\mathfrak{h}\otimes
\mathfrak{h}\subset U\mathfrak{h}$.} Therefore, we can view a Langevin process
as a left-invariant random walk in the group ${\rm Diff}(X)$.

Random processes, whose Euler (Hamilton) representation is described by
Eq.~(\ref{FP-general}), have been considered for the case of a
finite-dimensional Lie group $H$, with the corresponding finite-dimensional
$\mathfrak{a}(H)$, in the context of the Lyapunov exponent statistics in
chaotic systems for $H=SL(n;\mathbb{R})$ with the corresponding
$\mathfrak{a}(SL(n;\mathbb{R}))=sl(n;\mathbb{R})$. This includes the passive
scalar turbulence model \cite{CKL90th}, where chaotic behavior of the linear
infinitesimal deviations is modeled by stochastic behavior, as well as linear
\cite{RobMuz00} and nonlinear \cite{CM08b,CM08c} responses for weak-noise
$\kappa\to 0$ geodesic flows in Riemann surfaces of genus $g\ge 2$ with
constant negative curvature that corresponds to the case of
$H=SL(2;\mathbb{R})$.

It is the dynamical symmetry, expressed by Eq.~(\ref{FP-general}), i.e., the
fact that the relevant FP operator $\hat{{\cal L}}\in U\mathfrak{a}(H)$ belongs
to the universal enveloping algebra associated with the Lie algebra
$\mathfrak{a}(H)$ of the dynamical symmetry group $H$ that makes the
aforementioned problems to be tractable on the analytical 
level~\cite{CM08b, CM08c}, e.g., by
decomposing the space of distributions into irreducible representations and
making use of the fact that in each representation the evolution occurs
independently (which is due to the dynamical symmetry).

The stochastic equation (\ref{eq:Langevin-general}) should be properly
regularized. We choose a time regularization by representing the random vector
field $\xi(t)$ as a piece-wise constant function of time. Specifically we
split the time interval $[0,t]$ into short segments of length $\varepsilon$ and
assume the random vector field to be time-independent on each segment with the
correlation function
\begin{equation}
\label{Langevin-general-reg} \langle \xi^{j}(x) \rangle=0, \;\;\;
 \langle \xi^{i}(x)\xi^{j}(y) \rangle = 2\kappa \varepsilon^{-1} G^{ij}(x,y) \, ,
\end{equation}
with no correlations between the random vector fields on different segments.

Spatial regularization can be achieved by restricting the random field $\xi$ to
a finite-dimensional vector subspace ${\cal U}$ of the space
of smooth vector fields ${\rm Vect}(X)$ together with a positively defined
scalar product in ${\cal U}$, which naturally defines Gaussian fluctuations of
the allowed vector fields $\xi\in {\cal U}$. Obviously for a
regularized correlation function we have $G\in {\cal U}\otimes {\cal U}\in {\rm
Vect}(X)\otimes {\rm Vect}(X)$.

\subsection{Higher-Dimensional Stochastic Currents and Poincar\'e Duality}
\label{sec:currents}

Conventional empirical currents have been identified \cite{CCMT09} as
topologically protected observables. The approach is based on the notion of
cycles associated with stochastic trajectories and measurements. In this
manuscript, a $k$-dimensional cycle in $X$ is understood as a map (smooth,
continuous, or piece-wise smooth depending on the context) map $\gamma: K \to
X$ from a smooth oriented compact $k$-dimensional manifold $K$  to $X$. The
cycles can be added by taking a union, i.e., for $\gamma: K \to X$ and
$\gamma': K' \to X$ we define $\gamma+\gamma': K\sqcup K' \to X$. We also
define $-\gamma$ by just reversing the orientation of $K$. Therefore the cycles
form an abelian group. We call $\gamma\sim 0$ if there is a map (bordism)
$\chi: M\to X$ for a smooth $(k+1)$-dimensional manifold $M$ with the border
$\partial M=K$ so that its restriction to the border reproduces $\gamma$, i.e.,
$\chi|_{\partial M}=\gamma$. We call $\gamma\sim \gamma'$ when
$\gamma-\gamma'\sim 0$. The abelian group obtained from the group of
$k$-dimensional cycles by considering the equivalence classes $[\gamma]$ with
respect to the described above equivalence relation (referred to as bordism
equivalence) is called the {\it oriented bordism homology} group \cite{MS74} of
$X$ and will be denoted $H_{k}(X)$. In dealing with empirical currents we will
work with the {\it homology with real coefficients} defined as vector spaces
$H_{k}(X;\mathbb{R})=H_{k}(X)\otimes_{\mathbb{Z}}\mathbb{R}$.\footnote{The
  theory of higher-dimensional empirical currents can be also developed by
  understanding the cycles $\gamma$ as so-called singular cycles
  \cite{Spanier95}. In this case the currents are generated by stochastic
  motion of singular, rather than oriented bordism, cycles, and $H_{k}(X)$
  should be understood as {\it singular homology} \cite{Spanier95}. As we will
  see, the long-time statistics of the generated currents does not depend on the
  particular choice of the cycle type. This relies on the fact that the natural
  morphism from bordism homology to singular homology becomes an equivalence over
  real coefficients. For this reason we are also using the notation
  $H_{\bullet}(X)$ for oriented bordism homology, which is more common for
  singular case, instead of the standard $\Omega_{\bullet}(X)$.}

In the standard case of a stochastic trajectory represented by a map
$\eta:[0,t]\to X$, as argued in \cite{CCMT09}, the analysis can be restricted to
closed trajectories represented by $1$-cycles $\eta:S^{1}\to X$, so that the
equivalence class $t^{-1}[\eta]\in H_{1}(X;\mathbb{R})$ represents the
contribution of a stochastic trajectory $\eta$ to the empirical current
$\omega\in H_{1}(X;\mathbb{R})$. As outlined in \cite{CCMT09}, Poinc\'are
duality allows for an equivalent interpretation of the empirical current as a
set of components represented by a set of fluxes over a set of
$(m-1)$-dimensional cycles $\alpha:K \to X$, where the intersection index
$t^{-1}[\eta]\cdot[\alpha]$ is viewed as the contribution of the stochastic
trajectory to the component $\omega_{\alpha}\in\mathbb{R}$ of the empirical
current $\omega$.\footnote{In particular this means that the current components
  are defined as ${\omega}_{\alpha}=\omega\cdot[\alpha]$. We also reiterate
that the intersection index is well-defined on the level of equivalence
classes.}

In this manuscript we extend the notion of empirical currents to higher dimensions, i.e., we introduce currents generated in the full homology $H_{\bullet}(X;\mathbb{R})$ with $\bullet=1,2,\ldots,m$, so that the case $\bullet=1$ reproduces the standard empirical currents considered in \cite{CCMT09}. This generalization rests on the interpretation of a Langevin process as a random walk in ${\rm Diff}(X)$, as described in subsection~\ref{sec:Langevin-regular}. To put it on a formal grounds, we denote by $f(\xi) : [0,t]\to {\rm Diff}(X)$ a stochastic trajectory that starts at the identity diffeomorphism ${\rm id}_{X}\in{\rm Diff}(X)$ and is generated by the noise realization $\xi : [0,t]\to {\rm Vect}(X)$ according to the Langevin equation [Eq.~(\ref{eq:Langevin-general})]. Note that for the time regularization described in subsection~\ref{sec:Langevin-regular} the trajectory $f$ is continuous and piece-wise smooth.

Given a stochastic trajectory $f(\xi)$ we can associate with any
$n$-dimensional cycle $\gamma : N\to X$, considered as an initial condition, an
$(n+1)$-dimensional chain, represented by $\eta(\xi,\gamma): [0,t]\times N\to
X$, with $\eta(\xi,\gamma)(\tau,y)=f(\xi;\tau)(\gamma(y))$, which can be viewed
as a stochastic trajectory in the space $X^{N}$ of smooth maps $N\to X$.\footnote{This correspondence is actually due to a natural action ${\rm
Diff}(X)\times X^{N}\to X^{N}$ of the diffeomorphism group in the space of
cycles.} If this trajectory were periodic (with period $t$), it would be
represented by a cycle $\eta(\xi,\gamma) : S^{1}\times N\to X$, whose homology
class $t^{-1}[\eta]\in H_{n+1}(X;\mathbb{R})$ will represent the contribution
of the stochastic trajectory $\eta$ to the higher-dimensional current
$\omega\in H_{n+1}(X;\mathbb{R})$, in full analogy with the conventional $n=0$
case. In the latter case, a stochastic trajectory, which generally has a boundary,
can be turned into a closed trajectory by connecting its ends with a geodesic
line, which does not affect the long-time asymptotic of the current
distributions~\cite{CCMT09}. A generalization of such closing procedure in the
higher-dimensional case does not look straightforward. It will be addressed in
some detail in subsection~\ref{sec:generating-functions}. Here we will present
an intuitive picture based on the notion of empirical current components.

To this end, we introduce a $k$-dimensional cycle $\alpha:K\to X$ with
$k=m-(n+1)$ and introduce the intersection index $\eta\cdot\alpha$. An
attempt can be made in a standard way: we slightly perturb the piece-wise
smooth trajectory $\eta:[0,t]\to X^{N}$ to obtain a smooth map
$\tilde{\eta}:[0,t]\times N \to X$, we further slightly perturb $\tilde{\eta}$
and $\alpha$ to achieve their transversal intersection in a finite number of
points. We define the intersection index $\eta\cdot\alpha$ in a standard
transversality-based way as a sum over the intersection points weighted with
the proper $\pm 1$ sign factors determined by the orientations. If the
trajectory $\eta$ was periodic, then $\eta$ would represent an
$(n+1)$-dimensional cycle and the intersection index $\eta\cdot\alpha$ would be
well defined, in particular it would not depend on the adjustments needed to
achieve trasnversality, as described above. Later in the manuscript we will
show that in the realistic case of non-periodic trajectories the dependence on
the adjustments vanishes statistically in the long-time $t\to\infty$ limit,
which implies that the intersection index is well-defined statistically in the
relevant for us $t\to\infty$ limit. Such a situation will be also described as
that statistically the path integration is restricted to almost closed
trajectories. Having said that, to set the stage we just pretend that
$\eta\cdot\alpha$ is a well-defined quantity. Under such an assumption we can
come up with a simple and intuitive path-integral expression for the generating
function $Z(\lambda_{\alpha})$ that fully characterizes the probability
distribution function for the $\omega_{\alpha}$ component of the
higher-dimensional current $\omega\in H_{n+1}(X;\mathbb{R})$:
\begin{equation}
\label{Z-of-lambda} Z(\lambda_{\alpha};t) = \int{\cal D}\xi e^{-\kappa^{-1} S(\xi)+\lambda_{\alpha}\eta(\xi,\gamma)\cdot\alpha},
\end{equation}
where $S(\xi)$ is a quadratic in $\xi$ action that produces the correlation functions of the random vector field, given by Eq.~(\ref{eq:Langevin-general}), and the normalization factor is assumed to be absorbed in the measure ${\cal D}\xi$. Note that the path integral in Eq.~(\ref{Z-of-lambda}) should be regularized as described in subsection~\ref{sec:Langevin-regular}.

The above arguments can be formalized as follows. Denote by $\omega(\gamma,
\alpha; t)$ the average value of the flux through the cross section cycle
$\alpha: K \to X$ produced over time by stochastic trajectories that start with
the cycle $\gamma: N \to X$. Averaging can be in principle performed using the
probability distribution function of the flux that corresponds to the
generating function, given by Eq.~(\ref{Z-of-lambda}). Let $k = {\rm dim}(N) +
1$, and $m = \dim(X)$. Then ${\rm dim}(K) = m- k$. Verify that the following
holds: for $t \to \infty$ the average flux does not depend on $t$ and depends
on $\gamma$ and $\alpha$ through their homology classes $[\gamma]$ and
$[\alpha]$, respectively. This creates a bilinear map $J: H_{k- 1}(X) \otimes
H_{m- k}(X) \to \mathbb{R}$. On the other hand the intersection index generates
a bilinear map ${\rm Int}: H_{k}(X) \otimes H_{m- k}(X) \to \mathbb{Z}$, which
is known to be a non-degenerate over $\mathbb{R}$. This is one of possible
formulations of Poincar\'e duality~\cite{Spanier95}. Due to non-degenerate
nature of the Poincar\'e pairing, there is a unique linear map $\omega_k :
H_{k-1}(X) \to H_k(X; \mathbb{R})$, hereafter referred to as the {\it flux map}
so that $J([\gamma] \otimes [\alpha]) = {\rm Int}(\omega_{k}([\gamma]) \otimes
[\alpha])$ for any $[\gamma]$ and $[\alpha]$.

\subsection{Examples of Higher-Dimensional Currents and Fluxes, Generated by Deterministic Flows}
\label{sec:higher-currents-intro}
As explained in some detail in the previous subsection, a higher-dimensional
stochastic flux is a new observable associated with a Langevin process, given
by the intersection index of a higher-dimensional trajectory, produced by
motion of a higher-dimensional cycle, with a cross section represented by
another cycle of complimentary dimension to the trajectory dimension.
Since a stochastic trajectory is in fact a deterministic trajectory that
corresponds to some given realization of a stochastic flow, to understand the
nature of such an observable one needs to understand how the flux is generated
in the case of a deterministic flow. Switching to a stochastic setting is
conceptually simple: one just needs to perform averaging over (stochastic)
realizations of the flow with a proper probability measure. In the standard
set-up of points moving to form $1$-dimensional trajectories, the picture of
the flux as an intersection index of a trajectory with a cross section is very
intuitive. In the higher-dimensional case the picture is much less intuitive.
Therefore, before we move further, we consider in this section some simple, yet
non-trivial examples of deterministic flows and study the higher dimensional
currents/fluxes, generated by the aforementioned flows.

Our first example deals with a simple flow in a torus $X = T^2 = S^1 \times S^1$. Using the natural coordinates $-\pi \le \theta_1, \theta_2 \le \pi$, the flow is defined by a differential equation
\begin{equation}
\label{flow-torus-2} \frac{d\theta_1}{dt} = u, \;\;\; \frac{d\theta_2}{dt} =0.
\end{equation}

If we monitor the motion of points we can choose the cross section to be a
$1$-dimensional cycle $\alpha: S^1 \to S^1 \times S^1$ with $\alpha_1(\theta) =
(\theta, 0)$ or $\alpha_{2}(\theta) = (0, \theta)$. A particle's trajectory
generated by the flow of Eq.~(\ref{flow-torus-2}) is periodic; during the
period $t = 2\pi/u$ it will not cross the cycle $\alpha_{1}$ and cross 
$\alpha_{2}$ once, producing the fluxes (intersection index per unit
time) $\omega_{1} = 0$, and $\omega_{2} = u(2\pi)$. Alternatively, we can represent the
flux associated with the flow as a $2$-dimensional vector $\bm{\omega} =
(\omega_1, \omega_2)$, with the dimension $2$ reflecting the two independent
$1$-dimensional cycles in the torus $T^2$. We can also look at the motion of
$1$-dimensional cycles $\gamma: S^1 \to S^1 \times S^1$, with $\gamma_{1} =
\alpha_1$ and $\gamma_2 = \alpha_2$. Since trajectories produced by motion of
$1$-cycles are $2$-dimensional, cross sections should be $0$-dimensional; so we
can choose a cross section to be a point, say $(\pi/2, \pi/2)$. The trajectory
that starts with $\gamma_2$ covers the whole torus during the time period $t=
2\pi/u$ producing one intersection, which corresponds to the flux
$\omega(\gamma_2) = u/2\pi$. The cycle $\gamma_1$ moves along itself, so that
there are no intersection and $\omega(\gamma_1) = 0$. This example clearly
demonstrates that the multidimensional flux depends on the moving cycle
$\gamma$, actually on its homology class $[\gamma]$. This can be summarized in
terms of the current/flux maps $\omega_{1}: H_0(X) \to H_{1}(X; \mathbb{R})$
and $\omega_{2}: H_1(X) \to H_{2}(X; \mathbb{R})$ as follows. Appreciating the
fact that 
\begin{equation}
  \begin{aligned}
H_0 (T^2) &= \mathbb{Z} \\
H_1 (T^2) &= \mathbb{Z} \oplus \mathbb{Z} \\
H_2 (T^2) &= \mathbb{Z} \, ,
\end{aligned}
\end{equation}
as well as $[\gamma_1] = 1 \oplus 0$ and
$[\gamma_2] = 0 \oplus 1$ form a basis set in $H_1(T^2) = \mathbb{Z} \oplus
\mathbb{Z}$, we can write $\omega_1(r) = ru(2\pi)^{-1} \oplus 0$, $\omega_{2}(r
\oplus s) = su(2\pi)^{-1}$, with $r, s \in \mathbb{Z}$.

Our next example is a flow in a $3$-dimensional torus $X = T^3 = S^1 \times S^1
\times S^1$. Using the natural coordinates $-\pi \le \theta_1, \theta_2,
\theta_3 \le \pi$, the flow is defined by a differential equation
\begin{equation}
\label{flow-torus-3} \frac{d\theta_1}{dt} = u, \;\;\; \frac{d\theta_2}{dt} =0, \;\;\; \frac{d\theta_3}{dt} =0
\end{equation}
This example can be analyzed in a manner, similar to how its low-dimensional
counterpart was studied above; therefore, details will be omitted. Consider for
$X$ the cycles $\mu : \{ * \} \to X$, $\gamma_{a} : S^1 \to X$,
$\alpha_{a} : S^1 \times S^1 \to X$, with $a = 1, 2, 3$, and $\nu : T^3 \to X$,
of dimensions $0$, $1$, $2$, and $3$, respectively, that form the basis set in
homology. Namely $\mu(*) = * \in X$, $\nu = {\rm id}_{X}$; $\gamma_a$ embeds
$S^1$ in $T^3$ along the $a$-th components of $S^1$ in $T^3$, whereas
$\alpha_a$ embeds $T^2$ in $T^3$, missing the $a$-th component. We have for
intersection indices $[\mu] \cdot [\nu] =1$, $[\gamma_a] \cdot [\alpha_b] =
\delta_{ab}$. The flux maps are non-trivial in all dimensions; after a simple
and transparent computation we arrive at
\begin{align}
  \begin{aligned}
\label{flux-map-torus-3} 
\omega_{1}(s[\mu]) &= s[\gamma_1], \\
\omega_2(r[\gamma_{1}] + l[\gamma_{2}] + s[\gamma_{3}]) &= u(2\pi)^{-1}(l[\alpha_3] + 
s[\alpha_2]), \\ 
\omega_3(r[\alpha_{1}] + l[\alpha_{2}] + s[\alpha_{3}]) &= ru(2\pi)^{-1}[\nu]. 
\end{aligned}
\end{align}

The examples considered in this section so far may create the misleading impression
that higher-dimensional currents do not provide additional information on the
driven nature of the underlying dynamics, since their properties are contained
in the standard $1$-dimensional currents, the latter describing motion of
points. Therefore, we consider a flow on a $5$-dimensional manifold $X = S^3
\times S^2$ (which is simply connected, i.e., does not have non-contractible
$1$-dimensional cycles, and thus does not have standard $1$-dimensional fluxes)
and demonstrate that the flow produces a $3$-dimensional current/flux,
generated by directed motion of a non-contractible $2$-dimensional cycle. First,
we will represent the $3$-dimensional sphere as $S^3 = SU(2)$, using a
standard assertion
\begin{equation}
\label{S3-to-SU2} g(n_{0}, \bm{n}) = n_{0}\sigma_{0} + i\bm{n} \cdot \bm{\sigma}, \;\;\; n_{0}^{2} + \bm{n}^{2} = 1,
\end{equation}
where $\sigma_{0}$ and $\bm{\sigma} = (\sigma_{1}, \sigma_{2}, \sigma_{3})$ are
the unit and Pauli $2 \times 2$ matrices, respectively. The flow on $X = SU(2)
\times S^2$ is determined by the following differential equation for $(g,
\bm{n}) \in SU(2) \times S^2$
\begin{equation}
\label{flow-SU2xS2} \frac{dg}{dt} = ik(\bm{n} \cdot \bm{\sigma})g, \;\;\; \frac{d\bm{n}}{dt} = 0,
\end{equation}
with $k > 0$ being some arbitrary rate constant. We consider motion of a
$2$-dimensional cycle $\gamma: S^2 \to SU(2) \times S^2$, defined by
$\gamma(\bm{n}) = (1, \bm{n})$ and considered as the initial condition, with $1
\in SU(2)$, being the unit element in $SU(2)$ represented by $\sigma_{0}$. The
flow (in this case deterministic), defined by Eq.~(\ref{S3-to-SU2}) produces a
trajectory $\eta: \mathbb{R} \times S^2 \to SU(2) \times S^2$. Since
Eq.~(\ref{S3-to-SU2}) can be easily solved analytically, we arrive at the
following explicit expression for the trajectory that starts with $\gamma$
\begin{equation}
\begin{aligned}
\label{flow-SU2xS2-trajectory} \eta(t, \bm{n}) &= (\exp(ikt (\bm{n} \cdot \bm{\sigma})), \bm{n}) \\
&= (\cos(kt) + i(\bm{n} \cdot \bm{\sigma})\sin(kt), \bm{n}).
\end{aligned}
\end{equation}

To compute the $3$-dimensional flux produced by the flow over time $t$, we
consider the intersection index of the trajectory $\eta$
with a cross section, represented by a $2$-dimensional cycle $\alpha: S^2 \to
SU(2) \times S^2$, defined by $\alpha(\bm{n}) = (i\sigma_{3}, \bm{n})$. To
simplify the analysis, note that the considered trajectory is periodic with
the period $2\pi$ with respect to the dimensionless time $\tau = kt$, and
consider the intersection of $\eta$ with $\alpha$ restricted to the time period
$-\pi \le \tau \le \pi$. Defining intersection points as pairs of points
$(\tau, \bm{n}) \in \mathbb{R} \times S^{2}$ and $\bm{n}' \in S^{2}$, so that
$\eta(\tau, \bm{n}) = \alpha (\bm{n}')$, the latter representing the
intersection condition, we find two intersection points for a time period, the
``north'' intersection $\tau = \pi/2$, $\bm{n} = \bm{n}' = (0, 0, 1)$ and the
``south'' one $\tau = -\pi/2$, $\bm{n} = \bm{n}' = (0, 0, -1)$. Both
intersections occur at the same point $i\sigma_3 \in SU(2)$. To compute the
intersection index we need to weight each intersection with the local
intersection index $\pm 1$, obtained in the following way. Choose orientations
on $\mathbb{R} \times S^{2}$, $S^{2}$, and $SU(2) \times S^{2}$. Consider
oriented basis sets $(\bm{e}_0, \bm{e}_1, \bm{e}_2)$ and $(\bm{e}'_1,
\bm{e}'_2)$ at the tangent spaces at $(\tau, \bm{n}) \in \mathbb{R} \times
S^{2}$ and $\bm{n} \in S^{2}$, respectively, that represent the intersection.
Applying the differentials $d\eta$ and $d\alpha$ to the above basis sets we
obtain $3 + 2 = 5$ vectors that form a basis set in the tangent space at the
intersection point $(i\sigma_3, \bm{n}) \in SU(2) \times S^{2}$. If the
obtained basis set has the same orientation as a one that describes the chosen
orientation on $SU(2) \times S^{2}$ the local index is set to $1$; if
otherwise, it is set to $-1$. A simple computation presented
in~\ref{sec:compute-local-intersect} shows that the local indices of both
intersection points are the same, so that the contribution of a time period to
the intersection index $[\eta] \cdot [\alpha]$ is $2$. This means that the
intersection index during time $t$ is approximately equal to $2kt/(2\pi)$.
\footnote{As briefly explained above approximately means the following. If the
  trajectory is periodic, i.e., in this case $kt = 2\pi n$, the equality is
  exact. Still for long times $kt \gg 2\pi$ the equality holds approximately,
  with the relative error $\sim (kt)^{-1}$, even if the trajectory is not
periodic, the latter being a typical case.} This corresponds to the value of
the $3$-dimensional flux $\omega = 2k/(2\pi)$.

This last example has an interpretation in terms of the flux map as follows. 
For $X = S^3 \times S^2$, we have 
\begin{equation}
  H_0(X) \cong H_2(X) \cong H_3(X) \cong H_5(X) \cong \mathbb{Z} \, ,
  \label{}
\end{equation}
with $H_2(X)$ and $H_3(X)$ generated by $[\alpha]$ and $[\gamma]$,
respectively. The only nontrivial flux map $\omega_3 : H_2(X) \to H_3(X)$ has a
form $\omega_3(s[\gamma]) = 2sk(2\pi)^{-1}[\alpha]$.

\subsection{Supersymmetric Fokker-Planck Equation: Fermions and Differential Forms}
\label{sec:SFP-ferm-diff-forms}

It has been outlined in \cite{T-NK04} that the Supersymmetric Fokker-Planck
(SFP) equation describes stochastic motion of higher-dimensional objects, such
as $n$-dimensional surfaces in the configuration space $X$. In this subsection,
we present a brief review of the SFP equation, connect the fermion states to
differential forms and discuss a particle-hole symmetry that, being formulated
using the language of differential forms, reproduces the Hodge star
transformation \cite{GH94} that can be viewed as the supersymmetric counterpart
of the Poinc\'are duality.

We reiterate that our configuration space $X$ is an oriented compact smooth
manifold of dimension ${\rm dim}\,X=m$ and start with introducing $n$-fermion
states $\psi^{(n)}$
\begin{equation}
\label{define-fermion-state} \psi^{(n)}(x,\Theta)=\psi_{i_{1}\cdots i_{n}}^{(n)}(x)\Theta^{i_{1}}\cdots\Theta^{i_{n}},
\end{equation}
where $\Theta^{j}$ with $j=1,\ldots,m$ are Grassmann anticommuting variables,
i.e. $\Theta^{j}\Theta^{i}=-\Theta^{i}\Theta^{j}$, and therefore
$(\Theta^{i})^{2}=0$. The vector space of $n$-fermion states is denoted by
$A^{n}(X)$. Obviously, $A^{n}(X)\cong 0$ for $n>m$. Combining fermion states
with all possible numbers of fermions we obtain a graded supercommutative
algebra \cite{M97} denoted by $A^{\bullet}(X)$, whose elements (states)
\begin{equation}
\label{define-superposition-state} 
\begin{aligned}
  \psi(x,\Theta)&=\sum_{n=1}^{m}\psi^{(n)}(x,\Theta) \\
  &=\sum_{n=1}^{m}\psi_{i_{1}\cdots i_{n}}^{(n)}(x)\Theta^{i_{1}}\cdots\Theta^{i_{n}}
\end{aligned}
\end{equation}
are represented by superpositions of $m$-fermion states.\footnote{Using formal
  language, by introducing the fermion superposition states we constructed a
  $\mathbb{Z}_{2}$-graded smooth manifold (often referred to as a
  supermanifold), associated with the cotangent bundle over its bosonic
  substrate $X$. In this context the superposition states in
  Eq.~(\ref{define-superposition-state}) are referred to as the functions in
  the supermanifold (see, e.g., \cite{M97} for the details).} We will also use
  the ``bullet'' as a dummy index variable $\bullet=0,1,\ldots,m$.

By replacing the Grassmann variables $\Theta^{i}$ with the differential symbols
$dx^{i}$ we can recast Eq.~(\ref{define-fermion-state}) as
\begin{align}
  \begin{aligned}
\label{fermion-state-diff-form} 
\psi^{(n)}(x,\Theta) &=\psi_{i_{1}\cdots i_{n}}^{(n)}(x)\Theta^{i_{1}}\cdots\Theta^{i_{n}} \\ 
&=\psi_{i_{1}\cdots i_{n}}^{(n)}(x)dx^{i_{1}}\wedge\cdots\wedge dx^{i_{n}} \\
&=\psi^{(n)}(x,dx),
\end{aligned}
\end{align}
where the wedge product symbol emphasizes the anticommutative property of 
multiplication of the differential symbols. Most importantly,
Eq.~(\ref{fermion-state-diff-form}) identifies $n$-fermion states as rank $n$
differential forms on $X$, and therefore the fermion states inherit all
properties of differential forms. The most relevant properties for us are:
external differential, pull-back construction, ability to integrate an
$n$-fermion sate over an $n$-cycle or $n$-chain, and Stokes' theorem.

Following \cite{T-NK04} we introduce an operator ${\cal Q}:A^{\bullet}(X)\to
A^{\bullet+1}(X)$ by ${\cal Q}=\Theta^{j}\partial_{j}$, where hereafter we use
the notation $\partial_{j}=\partial/\partial x^{j}$ and
$\bar{\partial}^{j}=\partial/\partial \Theta^{j}$ (acting from the left). In
the language of differential forms it is known as the exterior differential
operator $d:A^{\bullet}(X)\to A^{\bullet+1}(X)$. Obviously ${\cal Q}^{2}=0$. A
fermion state of the form ${\cal Q}\psi$ is called exact, and a state $\psi$ with
${\cal Q}\psi=0$ is called closed. Obviously an exact state is closed. Denoting
by $B^{\bullet}(X)$ and $Z^{\bullet}(X)$ the subspaces of the exact and closed
states, respectively, the latter relation reads $B^{\bullet}(X)\subset
Z^{\bullet}(X)\subset A^{\bullet}(X)$. The related quotient spaces $H_{{\rm
DR}}^{\bullet}(X)=Z^{\bullet}(X)/B^{\bullet}(X)$ are known as
the de Rham cohomology of $X$, and are finite-dimensional
in the relevant case of compact $X$~\cite{GH94}. For a closed state $\psi\in Z^{\bullet}(X)$ its image in $H_{{\rm
DR}}^{\bullet}(X)$ is denoted $[\psi]\in H_{{\rm DR}}^{\bullet}(X)$ and
referred to as a (de Rham) cohomology class. We also associate with a vector
field $\xi\in {\rm Vect}(X)$ an operator $i_{\xi}: A^{\bullet}(X)\to
A^{\bullet-1}(X)$, called the inner derivative, and defined by $i_{\xi} =
\xi^{j}\bar{\partial}_{j}$.

A pull-back construction associates with a smooth map $g:Y\to X$ a linear map
$g^{*}:A^{\bullet}(X)\to A^{\bullet}(Y)$ that ``pulls'' the fermion states on
$X$ back to form the corresponding states on $Y$. The pull-back that commutes
with the exterior differential, i.e., $g^{*}{\cal Q}_{X}={\cal Q}_{Y}g^{*}$ 
so the rank is presevered, is
defined as
\begin{equation}
\label{define-pull-back}
\begin{split}
  (g^{*}\psi^{(n)})_{j_{1}\cdots j_{n}}(y)&\vartheta^{j_{1}}\cdots\vartheta^{j_{n}} \\ 
  =\psi_{i_{1}\cdots i_{n}}^{(n)}(g(y))&\frac{\partial g^{i_{1}}(y)}{\partial y^{j_{1}}}\vartheta^{j_{1}}\cdots\frac{\partial g^{i_{n}}(y)}{\partial y^{j_{n}}}\vartheta^{j_{n}},
\end{split}
\end{equation}
which reflects the following natural transformation from the variables
$(y,\vartheta)$ in $Y$ to the variables $(x,\Theta)$ in $X$.
\begin{equation}
\label{pull-back-cocrdinates} x=g(y), \;\;\; \Theta^{i}=(\partial g^{i}(y)/\partial y^{j})\vartheta^{j}.
\end{equation}

A fermion state $\psi\in A^{\bullet}(X)$ can be integrated over the manifold:
\begin{equation}
\label{define-integral-state} \int_{X}\psi=\int_{X}dx^{1}\ldots dx^{m}\int d\Theta^{1}\ldots d\Theta^{m}\psi(x,\Theta),
\end{equation}
with the integral over the Grassmann variables understood in the Berezin sense \cite{M97}. The Berezin integral is completely determined by the following set of rules:
\begin{equation}
\label{Berezin-integral} 
\begin{aligned}
\int d\Theta^{i} = 0, \qquad \qquad \int \Theta^{i}d\Theta^{i} &= 1, \\
\Theta^{i}d\Theta^{j} = - d\Theta^{j}\Theta^{i}, \,\,\,\quad d\Theta^{i}d\Theta^{j} &= - d\Theta^{j}d\Theta^{i}.
\end{aligned}
\end{equation}
The integral in the r.h.s. of Eq.~(\ref{define-integral-state}) is well
defined, since it does not depend on the choice of local coordinates (the
Jacobians that describe the transformations of the boson $dx$ and fermion
$d\Theta$ measures under a coordinate transformation cancel each other). Note
that only the maximal-fermion component $\psi^{(m)}(x,\Theta)$ of
$\psi(x,\Theta)$ contributes to the integral.

The above integration construction combined with the pull-back allows a fermion
state $\psi\in A^{\bullet}(X)$ to be integrated over a cycle or chain
$\gamma:N\to X$, that corresponds to the case of a manifold or manifold with a
border, respectively:
\begin{equation}
\label{define-integral-chain} \int_{\gamma}\psi=\int_{N}\gamma^{*}\psi,
\end{equation}
with the non-zero contribution provided by the $\psi^{(n)}$ component only.

Stokes' theorem in the case of smooth chains implies the
following: for a chain $\gamma:N\to X$ and fermion state $\psi\in A^{\bullet}(X)$ we have
\begin{equation}
\label{Stokes-theorem} \int_{\gamma}{\cal Q}\psi=\int_{\gamma|_{\partial N}}\psi.
\end{equation}

The Hodge star transformation \cite{GH94}, adopted to the language of fermion
states, $*:A^{\bullet}(X)\to A^{m-\bullet}(X)$ associates with an $n$-fermion
state $\psi^{(n)}$ an $(m-n)$-fermion state denoted by $*\psi$, and therefore
$*\psi$ provides a hole description of the original fermion state $\psi$. The
metric $g$ generates a standard scalar product of the states
\begin{align}
\label{sgandard-scalar-prod} 
(\psi^{(n)}, \varphi^{(n)}) &= \int_{X}\frac{dx}{\sqrt{g(x)}}n!g^{i_{1}j_{1}}(x) 
\cdots g^{i_{n}j_{n}}(x)\psi_{i_{1} \cdots i_{n}}^{(n)}\varphi_{j_{1} \cdots j_{n}}^{(n)} \nonumber \\
&= \int \psi^{(n)} (*\varphi^{(n)}),
\end{align}
with the second equality being in fact the defining property of the Hodge star operator.

The Supersymmetric Fokker-Planck (SFP) operator can be presented in a very
simple form in the case of an arbitrary metric $g^{ij}(x)$ in the following way.
We start with introducing the force $f_{i}(x; t) = g_{ij}(x)u^{j}(x; t)$ that
corresponds to the velocity field in the Langevin equation, and further its
supersymmetric counterpart $F(x, \Theta; t) = f_{j}(x; t)\Theta^{j}$. The SFP
operator has a form
\begin{equation}
\label{define-SFP-operator} {\cal L} = \kappa({\cal Q}{\cal Q}_{F}^{\dagger} + {\cal Q}_{F}^{\dagger}{\cal Q}), \;\;\; {\cal Q}_{F} = {\cal Q} + \kappa^{-1}F
\end{equation}
with Hermitian conjugation defined using an obvious scalar product in the space
of states with given fermion number (or equivalently differential forms of
given rank). This is the only source of dependence of the operator ${\cal L}$
on the metric. Several comments are in place. First, the operator ${\cal L}$
preserves the fermion number, also referred to as the degree, and hence so
does the evolution governed by the SFP equation. Second, $[{\cal Q}, {\cal
L}]$, and therefore the SFP equation preserves closed and exact states. This
allows the dynamics to be restricted to closed states, with the SFP equation
written in a form
\begin{equation}
\label{SFP-eq-continuity} \frac{\partial\varrho}{\partial t} = -{\cal Q}J, \;\;\; J = -\kappa{\cal Q}_{F}^{\dagger}\varrho
\end{equation}
of a continuity condition, as it happens in the case of the standard FP
equation. One might think about the higher-dimensional counterpart,
defined by Eq.~(\ref{SFP-eq-continuity}), as a suitable candidate for an
average (over the stochastic process) higher-dimensional current density, in
terms of which the average fluxes can be expressed. The simple relations
between the higher-dimensional fluxes and current densities will be stated in
section~\ref{sec:higher-flux-driven}, with the derivations briefly sketched in
section~\ref{sec:Langevin-current-supersymmetry}. Finally we note that in a
particular case $F = 0$, the SFP operator in Eq.~(\ref{define-SFP-operator})
adopts the form of the Hodge-Laplace operator, widely used in Hodge theory,
whereas in a particular case of flat metric in $\mathbb{R}^m$ and potential
force $f_{j} = -\partial_{j}V$, the operator in Eq.~(\ref{SFP-eq-continuity})
reduces to the SFP, considered in~\cite{T-NK04}.

\subsection{Poincar\'e Duality and Higher-Dimensional Fluxes Generated in Stationary and Periodically Driven Systems}
\label{sec:higher-flux-driven}

The relation between the higher-dimensional current densities
[Eq.~(\ref{SFP-eq-continuity})] and generated fluxes is naturally formulated in
terms of a version of Poincar\'e duality that involves de Rham
cohomology~\cite{GH94}, defined in section~\ref{sec:SFP-ferm-diff-forms} which
can be formulated as follows. Poincar\'e duality consists of the set of isomorphisms
\begin{equation}
\label{poincare-de-rham} H_k(X;\mathbb{R}) \cong H_{{\rm DR}}^{m-k}(X), \;\;\; k = 0, \ldots, m,
\end{equation}
naturally defined using the two pairings
\begin{equation}
\label{kronecker-dual} 
  \begin{aligned}
\langle \cdot \; , \; \cdot \rangle &: H_k(X;\mathbb{R}) \otimes H_{{\rm DR}}^k(X) \to \mathbb{R}, \\
 \langle [\gamma] \; , \; [\omega] \rangle &= \int_{N}\gamma^{*}(\omega), \;\;\; \gamma: N \to X, 
 \end{aligned}
\end{equation}
and
\begin{equation}
 \begin{aligned}
 \langle \cdot \; , \; \cdot \rangle &: H_{{\rm DR}}^{m-k}(X) \otimes H_{{\rm DR}}^k(X) \to \mathbb{R}, \\
 \langle [\psi] \; , \; [\omega] \rangle &= \int_{X} \psi \omega, \; \omega \in Z^k(X), \; \psi \in Z^{m-k}(X).
 \end{aligned}
\end{equation}

We will show in section~\ref{sec:Langevin-current-supersymmetry} that a
probability distribution ${\cal P}(\gamma)$ of $(k-1)$-dimensional cycles
$\gamma: N \to X$ allows a reduced description in terms of a closed state
$\varrho \in Z^{m-(k-1)}(X)$, with $[\varrho] \in H_{{\rm DR}}^{m-(k-1)}$
corresponding to $[\gamma] \in H_{k-1}(X)$ via Poincar\'e duality
[Eq.~(\ref{poincare-de-rham})], that satisfies the SFP equation
[Eq.~(\ref{SFP-eq-continuity})]. Let $\alpha: K \to X$ be an
$(m-k)$-dimensional cross-section cycle. It will be shown that the average flux
$\omega(\alpha; t)$ through the cross section $\alpha$, understood as the
intersection index of a stochastic trajectory with $\alpha$ per unit time,
averaged over the stochastic process can be represented in a form
\begin{equation}
\label{flux-finite-t} 
\omega(\alpha; t) = -\kappa t^{-1}\int_{0}^{t}d\tau\int_{K}\alpha^{*}{\cal Q}_{F(\tau)}^{\dagger}\varrho(\tau).
\end{equation}

The expression for the flux adopts a specially simple form for the long-time
limit in the cases of stationary and periodic driving. We start with a
stationary case, when $F(\tau) = F$, so that at $t \to \infty$ we can replace
$\varrho(\tau)$ with the stationary solution of the SFP equation
$\varrho([\gamma])$ whose cohomology class $[\varrho]$ is Poincar\'e dual to the
homology class $[\gamma]$ of the cycle $\gamma$ that participates in stochastic
dynamics. The average flux becomes time-independent and can be represented in the
form
\begin{equation}
\label{flux-stationary} 
\begin{aligned}
  \omega(\alpha; [\gamma]) &= \int_{K}\alpha^{*}\omega([\gamma]), \\
  \omega([\gamma]) &=  -\kappa{\cal Q}_{F}^{\dagger}\varrho([\gamma]), \\ 
  {\cal Q}\omega &= 0 \, .
\end{aligned}
\end{equation}
The last condition implies $\omega([\gamma]) \in Z^{m-k}(X)$ is closed, so
that taking the cohomology class $[\omega]$ defines a flux map

\begin{equation}
\label{flux-stationary-2} \omega: H_{k-1}(X) \to H_k(X; \mathbb{R}),
\end{equation}
which sends the homology class $[\gamma]$ of the moving cycle $\gamma$ to the
homology class dual to the cohomology class $[\omega([\gamma])]$ via Poincar\'e
duality [Eq.~(\ref{poincare-de-rham})]. It is important to note that the
long-time limit of the average flux turns out to be represented by a closed
state $\omega([\gamma])$, which means that the flux through a cross section
$\alpha$ depends on its homology class $[\alpha]$ only, as if the trajectories
were closed. As briefly discussed at the end of subsection~\ref{sec:currents},
this is interpreted as at the long-time limit the stochastic trajectories are
statistically closed.

In the periodic driving case, following~\cite{CS09}, we consider the driving
force to be represented by a potential $f_{j}(t) = -\partial_{j}V(t)$ that
depends on time periodically $V(t+T) = V(t)$. In the periodic driving case it
is advantageous to define the flux as the intersection index per one diving
protocol, rather than per unit time. The expression for the average flux
$\omega(\alpha, [\gamma])$ has the form of Eq.~(\ref{flux-stationary}) with the
integrand $\omega([\gamma])$ replaced with
\begin{equation}
\label{flux-periodic} \omega([\gamma]) =  \int_{0}^{T}d\tau(-\kappa{\cal Q}_{F}^{\dagger})\varrho([\gamma]; \tau),
\end{equation}
with $\varrho([\gamma]; \tau)$ being the unique closed solution of the SFP equation [Eq.~(\ref{SFP-eq-continuity})] with $F = -{\cal Q}V$ and $[\varrho([\gamma]; \tau)] = [\gamma]$.

In section~\ref{sec:Langevin-current-supersymmetry}, we present a derivation
that starts with Eq.~(\ref{flux-periodic}) and allows the average flux to be
expressed in terms of a higher-dimensional version of the Kirchhoff problem in
the continuous (rather than discrete) setting. The aforementioned Kirchhoff
problem can be formulated as follows. We first note that we have an onto
linear map ${\cal Q}: A^{m-k}(X) \to B^{m-k+1}(X)$, by definition.
An attempt to invert it faces the problem of ambiguity, since the map
has a kernel, which consists of the subspace $Z^{m-k}(X)$ of closed states. The
uncertainty can be fixed by requiring that the image of the inverse map is
orthogonal to the kernel. The above requirement allows to define a unique
inverse, known in linear algebra as {\it pseudo-inverse}, as long as a scalar
product in the space $A^{m-k}(X)$ of $(m-k)$-fermion states is chosen.

Consider a modified scalar product in the space $A^{m-k}(X)$, defined by
\begin{equation}
\label{scalar-prod-mod} (\alpha, \beta)_{V, \kappa} = (e^{\kappa^{-1}V}\alpha, \beta) = (\alpha, e^{\kappa^{-1}V}\beta),
\end{equation}
and let $A_{V, \kappa}$ be the pseudo-inverse of ${\cal Q}$ with respect to the modified scalar product [Eq.~(\ref{scalar-prod-mod})], i.e., it is completely identified by its properties
\begin{equation}
\label{pseudo-inv-cont} A_{V, \kappa}{\cal Q}\varphi = \varphi, \;\;\; (A_{V, \kappa}\varphi, \psi)_{V, \kappa} = 0, 
\end{equation}
for any $\varphi \in B^{m-k+1}(X)$, and any $\psi \in Z^{m-k}(X)$.
The problem of finding the pseudo-inverse in the above setting is hereafter referred to as the higher-dimensional continuous Kirchhoff problem.

In section~\ref{sec:Langevin-current-supersymmetry} we will derive the following expression for the average flux in terms of the pseudo-inverse of ${\cal Q}$
\begin{equation}
\label{flux-periodic-2} \omega([\gamma]) = -\int_{0}^{T}d\tau A_{V(\tau), \kappa}\dot{\varrho}([\gamma]; \tau).
\end{equation}
where $\dot{\varrho} = d\varrho/d\tau$ denotes the time derivative. Note that $\dot{\varrho} = {\cal L}\varrho = {\cal Q}(\kappa{\cal Q}_{-{\cal Q}V}^{\dagger})$ is closed, so that applying the pseudo-inverse $A_{V, \kappa}$ to $\dot{\varrho}$ in the r.h.s. of Eq.~(\ref{flux-periodic-2}) is legitimate. Also note that applying ${\cal Q}$ to Eq.~(\ref{flux-periodic-2}) yields, due to the first property in Eq.~(\ref{pseudo-inv-cont}), in the r.h.s. just the time integral of $\dot{\varrho}$ over a period, which is zero. Therefore, $\omega([\gamma])$ is exact, so that the same arguments, as presented earlier in the context of stationary driving, bring us to a well-defined flux map of the same form of Eq.~(\ref{flux-stationary-2}).

Finally we note that in the case of slow driving (adiabatic limit) the periodic solution of the SFP equation is well approximated $\varrho([\gamma]; \tau) = \varrho_{{\rm B}}([\gamma]; V(\tau), \kappa)$, with $\varrho_{{\rm B}}([\gamma]; V, \kappa)$ hereafter, with some minimal abuse of terminology, being referred to as a higher-dimensional Boltzmann distribution. It is formally defined as a unique closed stationary solution of the SFP equation with $F = -{\cal Q}V$ and $[\varrho_{{\rm B}}([\gamma])]$ being Poincar\'e dual to $[\gamma]$. So in the adiabatic limit we have explicitly
\begin{equation}
\label{flux-periodic-3} \omega([\gamma]) = -\int_{0}^{T}d\tau A_{V(\tau), \kappa}\frac{d}{d\tau}\varrho_{{\rm B}}([\gamma]; V(\tau), \kappa).
\end{equation}

\section{Langevin Processes, Higher-Dimensional Currents, and Supersymmetric Fokker-Planck Dynamics}
\label{sec:Langevin-current-supersymmetry}

In this section, we develop an approach that allows us to derive the
expressions [Eqs.~(\ref{flux-finite-t}), (\ref{flux-stationary}), and
(\ref{flux-periodic-2})] for the average higher-dimensional current in terms of
the boson/fermion states, introduced in section~\ref{sec:SFP-ferm-diff-forms},
and their evolution via the SFP equation. Our approach is based on introducing
the {\em reduced probability distributions} or {\em reduced measures} that are
much simpler objects compared to probability measures in the
infinite-dimensional spaces of cycles. These evolve according to the SFP equation
and contain all relevant information about the average higher-dimensional
fluxes.

\subsection{Generating Functions for Empirical Current Distributions}
\label{sec:generating-functions}

Probability distributions are efficiently studied by considering the associated
generating functions. It is convenient to replace the generating
function, given by Eq.~(\ref{Z-of-lambda}), with a more general $Z({\cal
A};\gamma,t)$, whose argument is an $k$-fermion state ${\cal A}\in A^{k}(X)$
\begin{align}
\label{eq:Z-of-A} 
  Z({\cal A};\gamma,t) &= \left\langle \exp\left(\int_{N\times[0,t]}\eta^{*}(\xi,\gamma){\cal A}\right) \right\rangle_{\eta} \\
  &= \int{\cal D}\xi e^{-\kappa^{-1} S(\xi)}\exp\left(\int_{N\times[0,t]}\eta^{*}(\xi,\gamma){\cal A}\right), \nonumber
\end{align}
where, same as in Eq.~(\ref{Z-of-lambda}), $S(\xi)$ is a quadratic function of
the random vector field that produces the proper correlation functions of the
latter [Eq.~(\ref{eq:Langevin-general})], and the generating function naturally
depends on the initial cycle $\gamma$.\footnote{Also note that the expression
  in the second exponent in Eq.~(\ref{eq:Z-of-A}) is just a properly defined
  integral $\int_{\eta}{\cal A}=\int_{N\times[0,t]}\eta^{*}{\cal A}$ of the
  $k$-fermion state ${\cal A}$ over the $k$-dimensional surface $\eta$.} We
  reiterate that according to the arguments presented above, the path-integral
  is properly and completely (i.e., in the time and configuration space
  domains) regularized.

Being focused on average currents we expand the generating function
[Eq.~(\ref{eq:Z-of-A})] to linear terms in its argument
\begin{align}
\label{Z-of-A-expand} 
Z({\cal A};\gamma,t) &= 1 + \int_{0}^{t}d\tau \int_{X}{\cal A}J(\tau; \gamma) + O({\cal A}^{2}) \\ 
&= 1 + \int{\cal D}\xi e^{-\kappa^{-1} S(\xi)}\int_{N\times[0,t]}\eta^{*}(\xi,\gamma){\cal A} + O({\cal A}^{2}), \nonumber
\end{align}
where $J(\tau; \gamma) \in A^{m-k}(X)$ is just a functional coefficient in the
above linear expansion, represented by a time-dependent $(m-k)$-fermion state
that also depends parameterically on the initial cycle $\gamma$.

If the stochastic trajectories were closed (periodic in time), i.e., the
corresponding smooth maps $\eta:N\times[0,t]\to X$ could be considered as
cycles $N\times S^{1}\to X$, then replacing ${\cal A}$ with ${\cal A}+{\cal Q}{\cal
B}$, referred to as a gauge transformation of ${\cal A}$, would not change the
${\cal A}$-dependent exponential term in Eq.~(\ref{eq:Z-of-A}) due to the
Stokes' theorem (see subsection~\ref{sec:SFP-ferm-diff-forms}) and the
generating function would be gauge invariant. Therefore, it makes sense to
refer to a situation when the generating function becomes gauge-invariant,
i.e., $Z({\cal A}+{\cal Q}{\cal B})=Z({\cal A})$ at long times as ``at long
times the trajectories are closed statistically''. The referred situation can
be described as follows. At long times the generating function has a large
deviation form $Z({\cal A};\gamma,t)\sim e^{-t{\cal F}({\cal A};[\gamma])}$
with ${\cal F}({\cal A};[\gamma])$ being gauge invariant and depending on the
initial cycle $\gamma$ via its homology class $[\gamma]$. In the
gauge-invariant case, the corresponding Cram\'er function ${\cal
S}(J;[\gamma])$, with $J\in A^{m-n-1}(X)$ is supported by the conserving ${\cal
Q}J=0$ currents and can be obtained from gauge-invariant ${\cal F}({\cal
A};[\gamma])$ by applying a standard Legendre transformation.

In the case when we restrict ourselves to closed ${\cal Q}{\cal A}=0$ fermion
states for the argument of the generating function, corresponding to
empirical currents $\omega=[J]$ rather than current densities,
the generating function depends on the cohomology class $[{\cal A}]\in
H^{k}(X)$. The corresponding Cram\'er function ${\cal S}(\omega;[\gamma])$ that
depends on $\omega=[J]\in H^{m-k}(X)\cong H_{k}(X;\mathbb{R})$, rather than the
current density $J\in Z^{m-k}(X)\subset A^{m-k}(X)$, can be obtained from
${\cal F}([{\cal A}];[\gamma])$ via a standard Legendre transformation.


It is worth to emphasize that, as opposed to the standard $n=0$ case, in higher
dimensions gauge invariance is not guaranteed, but is rather a statistical
property. Therefore, it can be broken, which means that the boundaries of the
stochastic trajectories $\eta$ start playing an important role. We will
demonstrate in this manuscript that gauge invariance is maintained at least on
the level of average values of current distributions. An interesting open
question is whether gauge invariance can be broken for large enough deviations
of the observed current $\omega$ from its stationary value. As a first step,
gauge symmetry breaking can be studied on the level of Markov-chain reduced
models described in the second manuscript.

Summarizing, to demonstrate the validity of the representation, given by
Eq.~(\ref{flux-finite-t}) it is enough to show that
\begin{equation}
  \begin{split}
\label{J-Lagrange-to-Euler} \int{\cal D}\xi e^{-\kappa^{-1} S(\xi)}\int_{N\times[0,t]}\eta^{*}(\xi,\gamma){\cal A} \\ = -\kappa \int_{0}^{t}d\tau\int_{X}{\cal A}{\cal Q}_{F(\tau)}^{\dagger}\varrho(\tau),
\end{split}
\end{equation}
where $\varrho(\tau)$ is the solution of the SFP equation. This will be done in
section~\ref{sec:SFP-currents} by introducing and studying the properties of
{\it reduced measures}.

\subsection{Reduced Measures, Supersymmetric Fokker-Planck Equation and Average Higher-Dimensional Currents}
\label{sec:SFP-currents}

We can interpret Eq.~(\ref{Z-of-A-expand}) as a path-integral representation of
the average current density $J$. However, path integrals are difficult to
calculate. A standard trick used for Langevin processes is to switch from the
Lagrangian, i.e., path-integral, picture to an equivalent, yet different
representation, often referred to as the Euler or Hamilton picture. This
transformation replaces a path integral by a deterministic linear equation for
the relevant distributions. We refer to the obtained equation as the twisted
Fokker-Planck (FP) equation, where the term ``twisted'' appreciates the fact
that the FP equation is modified by introducing the gauge field ${\cal A}$ that
serves as the argument of the generating function. Since in the
higher-dimensional $n>1$ case, the distributions ${\cal P}$ are defined in the
infinite-dimensional space $X^{N}$ of smooth maps $N\to X$, it is absolutely
imperative to view a distribution as an integration measure. That is, as a linear
functional that associates with a function $h:X^{N}\to \mathbb{R}$, a number
${\cal P}(h)$ referred to as the integral of $h$ with respect to the measure
${\cal P}$; a standard notation is often used:
\begin{equation}
\label{eq:inegral-measure} {\cal P}(h)=\int_{X^{N}}d{\cal P}(\zeta)h(\zeta).
\end{equation}

To perform the desired transformation to the Euler-Hamilton picture we make use
of the fact that a Langevin process can be viewed as a theory of stochastic
flows, as described in some detail in section~\ref{sec:Langevin-regular}. A
flow in $X$ generates a flow in the cycle space $X^{N}$ and further in the
space of measures. Denoting by ${\cal P}(\gamma; t)$ the value at time $t$ of
the measure, with the initial condition $P(\gamma; 0)(h) = h(\gamma)$, averaged
over stochastic flows (or equivalently vector fields) we arrive at the
following path-integral representation
\begin{equation}
\label{eq:define-P} {\cal P}(\gamma; t)(h) = \int{\cal D}\xi e^{-\kappa^{-1} S(\xi)}h(\eta(\xi,\gamma)(t)), \;\;\;
\end{equation}
where, according to our notation $\eta(\xi, \gamma)(t)\in X^{N}$ is the
$k$-cycle that represents the end-point of the stochastic trajectory $\eta(\xi,
\gamma)$ that starts at $\gamma\in X^{N}$ and is generated by the noise
realization $\xi$.

The family ${\cal P}(\gamma; t)$ of measures obviously satisfies the semi-group relation
\begin{equation}
\label{eq:P-group-relation} {\cal P}(\gamma, t+t')=\int_{X^{N}}d{\cal P}(\zeta; \gamma; t){\cal P}(\zeta; t'),
\end{equation}
which generally serves as a starting point for deriving the corresponding FP
equation by setting $t' = \varepsilon$, followed by implementing the limit
$\varepsilon \to 0$. However, as opposed to the standard $k = 1$ case, the
higher-dimensional situation brings an additional problem: The measure ${\cal
P}(\gamma; t)$, which obviously depends on the regularization, does not have a
well-defined $\varepsilon \to 0$ limit. This is easy to see, since the
regularized path integral in Eq.~(\ref{eq:define-P}) contains $({\rm dim}\,{\cal
U})^{t/\varepsilon}$ integrations and therefore the regularized ${\cal P}$ is
supported by a finite-dimensional surface in $X^{N}$, whose dimension $({\rm
dim}\,{\cal U})^{t/\varepsilon}$ grows rapidly with $\varepsilon \to 0$. We
address the problem by introducing the reduced measures that can be viewed as
restrictions of ${\cal P}$ to narrower classes of functions to be integrated.

The simplest reduction leads to simple reduced measures that allow average
currents to be handled, and, therefore, this is the only reduction that is
considered in this manuscript. Given a measure ${\cal P}$ in $X^{N}$,
understood in the sense of Eq.~(\ref{eq:inegral-measure}) we restrict it to a
vector subspace of functions $h_{\alpha}(\gamma)$, parameterized by
$(k-1)$-fermion states $\alpha \in A^{k-1}(X)$ of the form
\begin{equation}
\label{reduced-states} h_{\alpha}(\gamma) = \int_{\gamma}\alpha = \int_{N}\gamma^{*}\alpha.
\end{equation}
The reduced measure, associated with ${\cal P}$, is described by a closed $(n-
k+ 1)$-fermion state, denoted $\varrho({\cal P})$, completely described by the
following conditions
\begin{equation}
\label{define-reduction} 
{\cal Q}\varrho({\cal P}) = 0, \;\;\;\;\;\; \int_{X^{N}}d{\cal P}(\gamma)\int_{N}\gamma^{*}\alpha=\int_{X}\varrho({\cal P})\alpha ,
\end{equation}
for all $\alpha \in A^{k-1}(X)$.
Diffeomorphisms act on the states $\varrho$ by means of pull-backs, therefore,
flows act on these states in a well-defined way, and by the construction
[Eq.~(\ref{define-reduction})] the reduction commutes with the flow, i.e.,
\begin{equation}
\label{red-flow-commute} \varrho({\cal P}(t)) = \varrho({\cal P})(t),
\end{equation}
which simply means that the evolution of reduced measures can be obtained by
looking at the evolution of the corresponding representing super-states
$\varrho({\cal P})$.

Since vector fields can be viewed as infinitesimal diffeomorphisms, the action
of vector fields on the states $\varrho$ is easily identified and is known to
be given by the Lie derivative of $\varrho$ with respect to a vector field
$\eta$
\begin{equation}
\label{internal-derivative} L_{\eta}\varrho = {\cal Q}i_{\eta}\varrho + i_{\eta}{\cal Q}\varrho,
\end{equation}
with the inner derivative $i_{\eta}$ defined in
section~\ref{sec:SFP-ferm-diff-forms}. To identify the evolution of a
super-state we consider a short time interval $(t, t +\varepsilon)$, at which
the stochastic vector field is considered to be time-independent according to
our time regularization scheme, and neglect the time dependence of the velocity
field, which results in
\begin{equation}
\label{evolution-epsilon} 
\begin{aligned}
  \varrho(t + \varepsilon) &\approx \left\langle \exp\left(\varepsilon L_{u(t) + \xi}\right)\varrho(t) \right\rangle_{\xi} \\
  &\approx \varrho(t) + \varepsilon{\cal L}(t)\varrho(t) + O(\varepsilon^2),
\end{aligned}
\end{equation}
with the second approximate equality being just a definition of the evolution
operator ${\cal L}(t)$. Averaging over the random vector field is performed by
regularization of the space of allowed vector fields, restricting them to a
finite-dimensional space, spanned on a set of vector fields $(e_a|a = 1,
\ldots, N)$, i.e., representing the stochastic field in the form
\begin{equation}
\label{xi-regularize} 
\begin{aligned} 
  \xi^{j}(x) = \sum_{a = 1}^{n}\lambda_{a}e_{a}^{j}(x), \\ \langle \lambda_a \rangle = 0, \;\;\; 
  \langle \lambda_a \lambda_b \rangle = 2\kappa \varepsilon^{-1} \delta_{ab}, \\ G^{ij}(x, y) = \sum_{a=1}^{N}e_{a}^{i}(x)e_{a}^{j}(y)
\end{aligned}
\end{equation}
Expanding the expectation value in Eq.~(\ref{evolution-epsilon}) to first order
in $u$ and second order in $\xi$ (to keep all terms up to order $\epsilon$ after
averaging), substituting Eq.~(\ref{xi-regularize}) into
Eq.~(\ref{evolution-epsilon}) performing averaging explicitly, and comparing
the terms of first order in $\varepsilon$ we arrive at
\begin{equation}
\label{evolution-epsilon-2} {\cal L}(t)\varrho(t) = L_{u(t)}\varrho(t) + \kappa \sum_{a=1}^{N}L_{e_a}L_{e_a}\varrho(t).
\end{equation}
Recasting the Lie derivatives in Eq.~(\ref{evolution-epsilon-2}) in an explicit
form, and after some straightforward algebra we identify the operator ${\cal
L}$ with the SFP operator given by Eq.~(\ref{define-SFP-operator}).\footnote{To
  be precise, we should note that, similar to the standard FP operator case,
  the force field $F$, obtained in the above derivation contains additional
  terms, generated by the random field, which have the form of products of the
  fields $e_a$ and their gradients, and which can be expressed in terms of the
  covariant derivatives of $G^{ij}(x, y)$ with respect to $y$, taken at $y =
x$.} At this point we want to emphasize that the result, we have obtained, is
not only crucial in deriving Eq.~(\ref{J-Lagrange-to-Euler}), which will be
completed in the rest of this section, but also provides an important insight
on the SFP equation, interpreting the latter as the dynamical equation that
describes evolution of reduced measures, associated with stochastic motion of
higher-dimensional cycles under a Langevin process.

We complete this section with deriving Eq.~(\ref{J-Lagrange-to-Euler}). With
the machinery developed earlier in this section this becomes a simple and
straightforward task. Denoting by $I(t;\varepsilon)$ the contribution to the
l.h.s. of Eq.~(\ref{J-Lagrange-to-Euler}) that is due to the fragment of a
stochastic trajectory on the time interval $(t, t + \varepsilon)$, where,
according to the chosen time regularization, the stochastic field is
approximated by a time-independent one we obtain
\begin{align}
\label{J-L-to-E-epsilon} 
  I(t;\varepsilon) \nonumber \\ 
  &\approx \int_{0}^{\varepsilon}d\tau \left\langle\int_{X^N}d(\exp(\tau \pi(u(t) + \xi)) \right.
  {\cal P}(t)) \nonumber \\
  &\quad\quad\left.\times (\gamma)\int_{\gamma}i_{u(t) + \xi}{\cal A}\right\rangle_{\xi}   \\ 
  &= \int_{0}^{\varepsilon}d\tau \int_{X} \langle\varrho(\exp(\tau \pi(u(t) + \xi)){\cal P}(t)) i_{u(t) + \xi}{\cal A}\rangle_{\xi} \nonumber \\
  &= \int_{0}^{\varepsilon}d\tau \int_{X} \langle \exp(\tau L_{u(t) + \xi})\varrho(t)i_{u(t) + \xi}{\cal A} \rangle_{\xi}. \nonumber
\end{align}
where $\pi(\eta)$ denotes the action of the vector field $\eta$ on the space of
measures on $X^N$. The first equality is due to the path-integral
representation for the measure evolution [Eq.~(\ref{eq:define-P})] and the
semigroup property [Eq.~(\ref{eq:P-group-relation})], the second equality
follows from the defining property of reduction [Eq.~(\ref{define-reduction})],
whereas the third equality reflects compatibility of the reduction with
stochastic evolution [Eq.~(\ref{red-flow-commute})]. Expanding the exponent in
the most r.h.s. of Eq.~(\ref{J-L-to-E-epsilon}) and further performing the
averaging in the same way it was done in deriving
Eq.~(\ref{evolution-epsilon-2}) we arrive at
\begin{equation}
\label{J-L-to-E-epsilon-2} 
\begin{aligned}
  I(t;\varepsilon) &\approx \varepsilon \int_{X}\langle\varrho(t)(i_{u(t)}{\cal A}) + (L_{\xi}\varrho) (i_{\xi}{\cal A})\rangle \\
  &= \varepsilon \int_{X}\varrho(t)(i_{u(t)}{\cal A}) + \frac{\varepsilon\kappa}{2} \sum_{a=1}^{N}\int_{X}(L_{e_a}\varrho) (i_{e_a}{\cal A}),
\end{aligned}
\end{equation}
and after some straightforward transformations we obtain
\begin{equation}
\label{J-L-to-E-epsilon-3} I(t;\varepsilon) \approx -\varepsilon\kappa \int_{X}{\cal A}{\cal Q}_{F(t)}^{\dagger}\varrho(t)
\end{equation}
Performing the summation of the contributions $I(t;\varepsilon)$, taking the
limit $\varepsilon \to 0$, combined with applying
Eq.~(\ref{J-L-to-E-epsilon-3}) immediately results in
Eq.~(\ref{J-Lagrange-to-Euler}).

\subsection{Explicit Expression for Higher-Dimensional Currents, Generated in a Stationary and Periodically-Driven System}
\label{sec:current-continuous-explicit}

The expression for the stationary case flux [Eq.~(\ref{flux-stationary})], as shown earlier in section~\ref{sec:higher-flux-driven} follows immediately from the general expression for the generated flux in the time-dependent case [Eq.~(\ref{flux-finite-t})]; the latter as derived in sections~\ref{sec:generating-functions} and \ref{sec:SFP-currents}. The expression for the periodic driving [Eq.~(\ref{flux-periodic})] also follows from Eq.~(\ref{flux-finite-t}) in a straightforward way. In this section we demonstrate how a more convenient expression for the flux [Eq.~(\ref{flux-periodic-2})] follows from Eq.~(\ref{flux-periodic}). In fact we will demonstrate a more general property, i.e. that the relation
\begin{equation}
\label{flux-relation} \int_{0}^{t}d\tau(-\kappa{\cal Q}_{-{\cal Q}V}^{\dagger})\varrho([\gamma]; \tau) = -\int_{0}^{t}d\tau A_{V(\tau), \kappa}\dot{\varrho}([\gamma]; \tau)
\end{equation}
holds in a general time-dependent case. This will be demonstrated by deriving the relation
\begin{equation}
\label{flux-relation-2} \kappa {\cal Q}_{-{\cal Q}V}^{\dagger}\varrho([\gamma]; \tau) = A_{V(\tau), \kappa}\dot{\varrho}([\gamma]; \tau).
\end{equation}

To that end we recast ${\cal Q}_{-{\cal Q}V} = e^{\kappa^{-1}V} {\cal Q} e^{-\kappa^{-1}V}$, which implies ${\cal Q}_{-{\cal Q}V}^{\dagger} = e^{-\kappa^{-1}V} {\cal Q}^{\dagger} e^{\kappa^{-1}V}$ and allows Eq.~(\ref{flux-relation-2}) to be recast in the form
\begin{equation}
\label{flux-relation-3} \kappa e^{-\kappa^{-1}V} {\cal Q}^{\dagger} e^{\kappa^{-1}V} \varrho([\gamma]; \tau) = A_{V(\tau), \kappa}\dot{\varrho}([\gamma]; \tau).
\end{equation}
Due to uniqueness of the pseudo-inverse (for a fixed scalar product), to show the validity Of Eq.~(\ref{flux-relation-3}) it is enough to verify that its l.h.s. satisfies the defining properties of the pseudo-inverse [Eq.~(\ref{pseudo-inv-cont})]. We have for the first property
\begin{equation}
\label{flux-relation-4-a} {\cal Q} \kappa e^{-\kappa^{-1}V} {\cal Q}^{\dagger} e^{\kappa^{-1}V} \varrho([\gamma]; \tau) = {\cal L}\varrho([\gamma]; \tau) = \dot{\varrho}([\gamma]; \tau)
\end{equation}
and for the second one
\begin{equation}
\label{flux-relation-4-b} 
\begin{aligned}
  (e^{-\kappa^{-1}V} {\cal Q}^{\dagger} e^{\kappa^{-1}V} \varrho, \psi)_{V, \kappa} &= ({\cal Q}^{\dagger} e^{\kappa^{-1}V} \varrho, \psi) \\
  &= (e^{\kappa^{-1}V} \varrho, {\cal Q}\psi) \\
  &= 0,
\end{aligned}
\end{equation}
since ${\cal Q}\psi = 0$, which completes the derivation.

\section{Discussion}
\label{sec:discusssion}

In this manuscript, we have extended the concept of currents and fluxes
generated in non-equilibrium (driven) stochastic processes to higher
dimensions. We have done this in the
continuous case, where the higher-dimensional currents characterize the same process as
standard stochastic currents, i.e., Langevin stochastic dynamics on a
manifold $X$ (of arbitrary dimension ${\rm dim}(X) = m$) with inhomogeneous
noise. This has been achieved by applying the following key steps.

(i) We considered a Langevin process the way it should be considered, i.e., as
a process that involves deterministic, as well as random components of the
velocity field, referred to as $u^j(x, t)$ and $\xi^j(x, t)$, respectively.
This is a natural view if one interprets a Langevin process as a result of
eliminating fast environmental (bath) degrees of freedom, so that the random
component of the velocity field is generated by the force field that describes
the system-bath interaction, and further applying the overdamped limit. By
introducing the more general quantity $G^{ij}(x, y)$ that
characterizes the correlations of the random field $\xi^j(x, t)$ at different
points, rather than its reduced counterpart $g^{ij}(x) = G^{ij}(x, x)$ that is
usually involved in a Langevin equation, we can view a Langevin processes as a
theory of stochastic flows or, equivalently, random walks in the space of
diffeomorphisms of the underlying manifold $X$.

(ii) Such interpretation of a Langevin process allows stochastic dynamics of
higher-dimensional (extended) objects to be considered; in particular one can
look at what happens with $(k-1)$-dimensional cycles, for $k = 0, \ldots, m-1$
under the random flow. Precisely, one can associate observables with
$k$-dimensional trajectories spanned as a result of motion of $(k-1)$-cycles.
Using an interpretation of the standard flux through a cross section as an
intersection index of a $1$-dimensional stochastic trajectory with an $(m-1)$-dimensional
cross section, we provide a rigorous definition of higher-dimensional fluxes as
new observables associated with a ``good old'' Langevin process, and also
reduce computation of their statistical properties to, sometimes non-trivial,
but still just technical details.

(iii) On the technical side, we generalized the Lagrangian (Langevin
equation) Euler-Hamiltonian (FP equation) correspondence to treat stochastic
dynamics of higher-dimensional cycles. This generalization faced some
technical difficulties, e.g., the infinite-dimensional nature of the space
$X^N$ of $(k-1)$-dimensional cycles, treated as smooth maps $\gamma : N \to X$,
with $N$ being a smooth compact $(k-1)$-dimensional manifold. The
aforementioned difficulties were bypassed by introducing a class of
probability measures $d{\cal P}(\gamma)$ on the infinite-dimensional cycle
space, understood as rules of integrations for functions $X^N \to \mathbb{R}$.
These measures, although well-defined, are objects too complex to be efficiently
handled. Therefore, by restricting the measures $d{\cal P}(\gamma)$ to narrower
subspaces of functionsx to be integrated, we introduce the so-called reduced
measures, represented by super-states $\varrho (x^1, \ldots, x^m; \Theta^1,
\ldots \Theta^m)$  on $X$, i.e., functions that depend on $m$ coordinates in
$X$, as well as $m$ Grassmann (anticommuting) variables. Such states has been
introduced in~\cite{T-NK04} in the context of supersymmetric stochastic theory
that describes stochastic motion of extended objects in a non-driven case.

(iv) Also on the technical side, we have derived an evolution equation for the
reduced measure $\varrho$ that turned out to be the SFP equation, presented
in~\cite{T-NK04} in the equilibrium (non-driven) case. The way it has been
derived is instructional. It is based on the aforementioned interpretation of a
Langevin process as a random walk in the space of diffeomorphisms. Therefore,
if we have any representation of the group of diffeomorphisms (or equivalently
the Lie algebra of vector fields, as its infinitesimal counterpart), i.e., a
vector space with a proper action of diffeomorphisms [the spaces of measures
$d{\cal P}(\gamma)$) and reduced measures $\varrho$ provide good and useful
examples], we can derive a FP equation in a standard fashion by applying the
short-time $\varepsilon \to 0$ evolution operator determined by the action of
the total vector field $u + \xi$, expand it to first-order and second-order
terms in $u$ and $\xi$, respectively with further averaging over the random
field, using its correlation function $G$. This means that the FP operator
exists as a universal object, and the FP equation for a given representation
can be obtained as an evaluation of the universal FP operator on a given
representation, as, e.g., in the case of reduced measures $\varrho$. The above
picture/derivation also implies an important property that the measure
reduction procedure commutes with stochastic evolution. This property allowed
us to derive a closed formal expression for the average value of a
higher-dimensional stochastic flux, generated over finite time, in terms of the
solution $\varrho(t)$ of the SFP equation. This completes the Lagrangian
Hamilton-Euler correspondence for higher-dimensional currents and fluxes.

(v) The aforementioned expression has been utilized to obtain analytical
expressions for the average flux in the cases of stationary and periodic
driving. We have introduced the higher-dimensional (supersymmetric) current
density operator and expressed the average flux as the integral over the cross
section [represented by a cycle $\alpha$ of  complimentary dimension $(n-k)$]
of the higher-dimensional (supersymmetric) current density $J$, obtained by
applying the current density operator to the unique stationary solution
$\varrho$ of the SFP equation, with the constraint that the cohomology class
$[\varrho]$ corresponds to the homology class $[\gamma]$ of the moving cycle
$\gamma$. We demonstrated that the flux depends only on the homology class
$[\alpha]$ that is interpreted as that in the long time limit the
$k$-dimensional trajectories are closed. We also showed that the average flux
depends on the initial value of the cycle $\gamma$ via its homology class
$[\gamma]$ only. Thus the higher-dimensional fluxes have been formulated in
terms that closely resemble the situation of standard currents. We derived
similar expressions for periodic driving, when the deterministic component of
the force has potential character, and depends on time in a periodic fashion,
the latter being the source of driving. The average flux is expressed in terms
of the periodic solution of the SFP equation and the solution of the
higher-dimensional continuous version of the Kirchhoff problem that can be
alternatively viewed as a pseudo-inverse operator to the supersymmetry operator
${\cal Q}$. This expression generalizes the result obtained in earlier work in
the context of periodic Markov chain processes on graphs to the
higher-dimensional continuous case~\cite{CKS12a, CKS13}. 

(vi) We identified the important role Poincar\'e duality plays in
higher-dimensional stochastic fluxes, the former being a basic and
celebrated concept in algebraic topology which establishes the equivalence of
homology and cohomology in complimentary dimensions. 
Homology appears in our
considerations in a straightforward way as equivalence classes $[\gamma]$,
$[\eta]$, and $[\alpha]$ of the initial cycle (whose homology class is stable
in stochastic evolution), stochastic trajectory, and cross section. The
cohomology appears in the de Rham form as the equivalence classes $[\varrho]$
and $[{\cal A}]$ of the reduced measures and argument of the generating
functions, respectively. It is not surprising that Poincar\'e duality appears
very naturally in dealing with fluxes, since fluxes are nothing more than
averaged intersection indices of stochastic trajectories with cross sections,
and an intersection index can be interpreted as an alternative view of
Poincar\'e duality.

In this manuscript, we focused on the average values of higher-dimensional
currents and fluxes. The question of the current and flux probability
distributions, which in the long time limit are well described by
large-deviation, also known as Cram\'er, functions ${\cal S}(J;[\gamma])$ and
${\cal S}(\omega; [\gamma])$~\cite{DZ09}. This is also known as the $2.5$-level 
theory~\cite{BC15}. The
problem can be treated by studying the generating function $Z({\cal
A};\gamma,t)$, defined by Eq.~(\ref{eq:Z-of-A}). In the case of standard
currents the path integral representation can be easily converted to the
Hamilton-Euler language, resulting in the so-called twisted FP equation, where
all spatial derivatives are just elongated, as it is done in gauge theories,
treating ${\cal A}$ (which in this case is a $1$-form, or a vector potential)
as a gauge field. This leads to a problem which is not substantially more
complex compared to finding the average currents. In the higher-dimensional
case the situation is quite different: one cannot obtain a closed equation of
the same level of complexity. The underlying reason is the extended nature of
higher-dimensional cycles; formally it appears as a fact that the reduced
measure $\varrho({\cal P})$ does not satisfy a closed equation. Our preliminary
analysis (these results will be published elsewhere) show that one can still
treat $\varrho({\cal P})$ as the first $p = 1$ terms in the infinite hierarchy
of reduced measures $\varrho_p({\cal P})$, described by super-states on the
Cartesian products $X^{\times p}$ of $p$ copies of $X$ and derive a set of SFP
equations for $\varrho_p$ that involve $\varrho_{p+s}$ with $s=0, 1, 2$.
However, it is promising that if one expands the generating function in powers of
${\cal A}$, which boils down to computing the higher moments of the flux, and
if one is interested in the moments up to the $p$-th level, the hierarchy can
be truncated on the $p$-th level. This implies that the higher moments can be
computed within the same concept at a cost of working with higher-dimensional
spaces. If the Cram\'er function is analytical in its argument, then knowing
all moments can reproduce it for moderate values of the flux. However, even its
analytical character is not established yet; whereas its behavior at very large
deviations remains a completely open question.


Another possible extension, which might sound more moderate, but seems to be no
less important, is to study average currents and their second moments in the
vanishing noise limit $\kappa \to 0$, and try to establish stable connections
between the statistical properties of the generated fluxes and the qualitative
nature of the underlying deterministic dynamics. Strongly chaotic, e.g., mixing
systems, are of special interest here. An interesting system that allows almost
analytical treatment is the low-noise limit of a geodesic flow on a Riemann
surface of genus $g \le 2$ with constant (negative) curvature~\cite{CM08c}. 
On the one hand,
the system can be treated efficiently by implementing dynamical symmetry and
decompositions in irreducible representations of the group $SO(2, 1)$. On the
other hand, its phase space, restricted to an energy shell has non-trivial
homology in all dimensions, thus non-trivial fluxes occur in all possible
dimensions $k = 1, 2, 3$.

\appendix

\section{Computation of Local Intersection Indices}
\label{sec:compute-local-intersect}

In this appendix we compute the local intersection indices for the example, considered in section~\ref{sec:higher-currents-intro}, of a $3$-dimensional current, generated by a deterministic flow in $X = SU(2) \times S^{2}$. We start by choosing orientations on $\mathbb{R} \times N = \mathbb{R} \times S^2$, $K = S^2$, and $X = SU(2) \times S^2$. We will be formulating everything in terms of dimensionless time $\tau = kt$. Orientations of cartesian products will be given by the orientations of their components. Orientation of $\mathbb{R}$ is given by a global basis set represented by a constant vector field, determined the unit vector $e_{0}(\tau) = e_0$, so that differentiation with respect to this vector field is given by $\partial/\partial\tau$. Orientation of $S^2$ is chosen by picking at point $\bm{n}$ any basis set $(\bm{e}_1, \bm{e}_2) = (\bm{e}, [\bm{n}, \bm{e}])$ with $(\bm{e} \cdot \bm{n} = 0$, and $\bm{e}^2 = 1$. Orientation of $SU(2)$ is chosen by introducing a global basis set $(\bm{u}_{1}(g), \bm{u}_{2}(g), \bm{u}_{2}(g))$, generated by right-invariant vector fields, i.e., we set $\bm{u}_{a}(g) = i\sigma_{a}g$, for $a = 1, 2, 3$. Note that in defining the basis sets in $S^2$ we made use of standard embeddings $S^2 \subset \mathbb{R}^3$ and $SU(2) \cong S^3 \subset \mathbb{R}^4$. We further note that, since the maps $\eta: \mathbb{R} \times S^{2} \to SU(2) \times S^2$ and $\alpha: S^2 \to SU(2) \times S^2$ are both the identity on their $S^2$ components, in comparing the basis sets we can restrict ourselves to considering the map $F: \mathbb{R} \times S^{2} \to SU(2)$ that represents the first components of the map $\eta$, defined by Eq.~(\ref{flow-SU2xS2-trajectory})
\begin{equation}
\label{define-F} 
\begin{aligned}
  F(\tau, \bm{n}) &= \exp(i\tau (\bm{n} \cdot \bm{\sigma})) \\ 
  &= \cos(\tau) + i(\bm{n} \cdot \bm{\sigma})\sin(\tau),
\end{aligned}
\end{equation}
and compare the basis sets $(dF(\tau, \bm{n})(e_0), dF(\tau, \bm{n})(\bm{e}),
dF(\tau, \bm{n})([\bm{n}, \bm{e}]))$ with $(\bm{u}_1, \bm{u}_2, \bm{u}_3)$.
The action of the differential $dF$ is easily identified by noting that
relaxing the condition $\bm{n}^2 = 1$ in Eq.~(\ref{define-F}) defines a map $F:
\mathbb{R} \times \mathbb{R}^3 \to \mathbb{R}^4$ that produces the original map
$F: \mathbb{R} \times S^{2} \to SU(2)$ by means of an obvious restriction,
resulting in
\begin{equation}
\label{dF-compute} 
\begin{aligned}
  dF(e_0) &= \frac{\partial F}{\partial\tau} = -\sin(\tau) + i(\bm{n} \cdot \bm{\sigma})\cos(\tau), \\ 
  dF(\bm{e}_{s})&= \frac{\partial F}{\partial\bm{n}} \cdot \bm{e}_{s} = i\bm{\sigma} \cdot \bm{e}_{s} \sin(\tau), \;\;\; s = 1,2
\end{aligned}
\end{equation}
To compare the basis set, represented by Eq.~(\ref{dF-compute}) with the
right-invariant basis set in $SU(2)$, we compute $dF(e_0)(F(\tau, \bm{n})^{-1},
dF(\bm{e})(F(\tau, \bm{n}))^{-1} \in su(2)$.  For the ``north'' intersection
point we have $\tau = \pi/2$, $\bm{n} = (0, 0, 1)$, $\bm{e}_1 = (1, 0, 0)$,
$\bm{e}_2 = (0, 1, 0)$. We find $F(\tau, \bm{n}) = i\sigma_3$, $\sin(\tau) =
1$, $\cos(\tau) = 0$ and further
\begin{equation}
\label{dF-compute-2} 
\begin{aligned}
  dF(e_0)F^{-1} &= -(i\sigma_3)^{-1} = i\sigma_3, \\  
  dF(\bm{e}_1)F^{-1} &= (i\sigma_1)(i\sigma_3)^{-1} = -\sigma_3\sigma_1 = -i\sigma_{2}, \\ 
  dF(\bm{e}_2)F^{-1} &= (i\sigma_2)(i\sigma_3)^{-1} = \sigma_2\sigma_3 = i\sigma_1,
\end{aligned}
\end{equation}
resulting in the basis set $(i\sigma_3, -i\sigma_2, i\sigma_1)$, which has the same orientation as $(i\sigma_1, i\sigma_2, i\sigma_3)$. The local intersection index is $+1$.

For the ``south'' intersection point we have $\tau = -\pi/2$, $\bm{n} = (0, 0, -1)$, $\bm{e}_1 = (1, 0, 0)$, $\bm{e}_2 = (0, -1, 0)$. We find $F(\tau, \bm{n}) = i\sigma_3$, $\sin(\tau) = -1$, $\cos(\tau) = 0$ and further
\begin{equation}
\label{dF-compute-3} 
\begin{aligned}
  dF(e_0)F^{-1} &= (i\sigma_3)^{-1} = -i\sigma_3, \\  
  dF(\bm{e}_1)F^{-1} &= -(i\sigma_1)(i\sigma_3)^{-1} = \sigma_3\sigma_1 = i\sigma_{2}, \\ 
  dF(\bm{e}_2)F^{-1} &= -(-i\sigma_2)(i\sigma_3)^{-1} = \sigma_2\sigma_3 = i\sigma_1,
\end{aligned}
\end{equation}
resulting in the basis set $(-i\sigma_3, i\sigma_2, i\sigma_1)$, which has the same orientation as $(i\sigma_1, i\sigma_2, i\sigma_3)$. The local intersection index is $+1$.

The obtained result can be also formulated in the following way. Due to periodicity of the map $F: \mathbb{R} \times S^2 \to SU(2)$ it defines a map $F: S^1 \times S^2 \to SU(2)$; our computation shows that the degree of the latter is equal to $+2$.

\section*{Acknowledgements}
This material is based upon work supported by the National Science Foundation under Grant No.\ CHE-1111350 and Simons Foundation Collaboration Grant No.\  317496


\end{document}